\documentclass[useAMS,usenatbib,usegraphicx]{mn2e}

\usepackage{times}

\newcommand{\JHK}{JHK_{\rm s}}
\newcommand{\Ks}{K_{\mathrm{s}}}
\newcommand{\AK}{A_{K_{\mathrm{s}}}}
\newcommand{\muGC}{\mu_0{({\mathrm{GC}})}}
\newcommand{\muLMC}{\mu_0{({\mathrm{LMC}})}}

\title[Cepheids and other variables near the GC]
{Cepheids and other short-period variables near the Galactic Centre}
\author[N. Matsunaga~{et~al.}]
{Noriyuki Matsunaga$^{1,2}$\thanks{E-mail:matsunaga@astron.s.u-tokyo.ac.jp},
Michael W. Feast$^{3,4}$,
Takahiro Kawadu$^{5}$,
Shogo Nishiyama$^{6}$,
\newauthor
Takahiro Nagayama$^{7}$,
Tetsuya Nagata$^{5}$,
Motohide Tamura$^{6}$,
Giuseppe Bono$^{8,9}$, 
\newauthor
and Naoto Kobayashi$^{10,2}$,
\\
$^{1}$ Department of Astronomy, School of Science, The University of Tokyo, 7-3-1 Hongo, Bunkyo-ku, Tokyo 113-0033, Japan \\
$^{2}$ Kiso Observatory, Institute of Astronomy, School of Science, The University of Tokyo, 10762-30, Mitake, Kiso-machi, Kiso-gun, Nagano 397-0101, Japan \\
$^{3}$ Astrophysics, Cosmology and Gravity Centre, Astronomy Department, University of Cape Town, Rondebosch, 7701, South Africa\\
$^{4}$ South African Astronomical Observatory, PO Box 9, Observatory 7935, South Africa \\
$^{5}$ Department of Astronomy, Kyoto University, Kitashirakawa-Oiwake-cho, Sakyo-ku, Kyoto 606-8502, Japan \\
$^{6}$ National Astronomical Observatory of Japan, 2-21-1 Osawa, Mitaka, Tokyo 181-8588, Japan \\
$^{7}$ Department of Astrophysics, Nagoya University, Furo-cho, Chikusa-ku, Nagoya 464-8602, Japan \\
$^{8}$ Dipartimento di Fisica, Universit\'{a} di Roma Tor Vergata, Via della Ricerca Scientifica 1, 00133 Rome, Italy \\
$^{9}$ INAF--Osservatorio Astronomico di Roma, Via Frascati 33, 00040 Monte Porzio Catone, Italy \\
$^{10}$ Institute of Astronomy, School of Science, The University of Tokyo, 2-21-1 Osawa, Mitaka, Tokyo 181-0015, Japan
}

\begin{document}

\date{Accepted 2012 November 1. Received 2012 October 31; in original form 2012 September 24}

\pagerange{\pageref{firstpage}--\pageref{lastpage}} \pubyear{2012}

\maketitle

\label{firstpage}

\begin{abstract}
We report the result of our near-infrared survey of 
short-period variable stars ($P < 60$~d)
in a field-of-view of $20^\prime \times 30^\prime$
towards the Galactic Centre.
Forty-five variables are discovered and we classify the variables
based on their light curve shapes and other evidence.
In addition to 3 classical Cepheids reported previously,
we find 16 type II Cepheids, 24 eclipsing binaries,
one pulsating star with $P=0.265$~d (RR~Lyr or $\delta$~Sct)
and one Cepheid-like variable whose nature is uncertain.
Eclipsing binaries are separated into
the foreground objects and those significantly obscured by
interstellar extinction. One of the reddened binaries 
contains an O-type supergiant and 
its light curve indicates an eccentric orbit.
We discuss the nature and distribution of type II Cepheids as well as
the distance to the Galactic Centre
based on these Cepheids and other distance indicators.
The estimates of $R_0$(GC) we obtained based on photometric data
agree with previous results obtained with kinematics of objects around the GC.
Furthermore, our result gives a support to the reddening law obtained
by Nishiyama and collaborators, $A_{\Ks}/E(H-\Ks)=1.44$,
because a different reddening law would
result in a rather different distance estimate.
\end{abstract}


\section{Introduction\label{sec:Intro}}

The Galactic Centre (hereafter GC) region is an important place
for many reasons.
A supermassive black hole exists in the direction of Sgr~A$^*$
within a complex region involving both hot and cold gas
(e.g.~\citealt{Morris-1996};
Genzel, Eisenhauer \& Gillessen, 2010). 
This region hosts the highest density of stars in the Galaxy, and
furthermore various stellar populations co-exist with different distribution
and characteristics
(Launhardt, Zylka \& Mezger, 2002). 
First, the extended Bulge with a scale of a few kilo-parsecs
has a triaxial or bar-like shape
(\citealt{Nakada-1991}; \citealt{Whitelock-1992}; \citealt{Stanek-1994})
and is populated predominantly by old stars
($\geq 10$~Gyr, \citealt{Zoccali-2003}; \citealt{Clarkson-2011}).
Secondly, the nuclear bulge show a disk-like distribution
with a radius $\sim 200$~parsecs, and a significant population of
young stars (a few Myr) are found in this region
(\citealt{Serabyn-1996}; \citealt{Figer-2004}; \citealt{YusefZadeh-2009}).
Finally, a dense stellar cluster with numerous massive stars
exist within a radius $\sim 10$~parsecs
(its core radius actually is much smaller,
$\sim 0.3$~parsec) around the central black hole \citep{Genzel-2003}.
The GC region provides us with a unique opportunity
to study not only stellar evolution but also phenomena in central parts of 
galaxies at close hand ($\sim 8$~kpc).
For instance, the most populous group of known young and massive stars, 
such as O-type stars and Wolf-Rayet stars, within the Galaxy exists there
(e.g.~\citealt{Mauerhan-2010}).

Pulsating variable stars are useful in studies of stellar populations.
In particular, Cepheids play important roles in a wide range of astronomy.
There are two groups of Cepheids, i.e.~classical Cepheids (hereafter CCEPs)
and type II Cepheids (T2Cs).
Both of them have period-luminosity relation (PLR), but the luminosities
at a given period differ by 1.5--2~mag (\citealt{Sandage-2006};
Matsunaga, Feast \& Soszy\'{n}ski, \citeyear{Matsunaga-2011a}).
CCEPs are pulsating supergiants with periods typically between 3 and 50~d,
evolved from intermediate- to high-mass stars (4--10~$M_\odot$).
On the other hand, T2Cs have similar periods to CCEPs, but 
are old and evolved from low-mass stars, $\sim 1~M_\odot$.
T2Cs are conventionally subdivided into the BL~Her and W~Vir stars
at periods less than 20 days and the RV~Tau stars with greater periods.
In addition, \citet{Soszynski-2008b}
identified peculiar W~Vir stars which tend to be brighter than the PLR
and to often show light curves with eclipsing or
ellipsoidal modulation.
There remain uncertainties in the properties and the evolution of T2Cs
(see the discussion in \citealt{Matsunaga-2011a}).

A serious difficulty in studying the stars towards the GC lies
in observing them beyond the severe interstellar extinction.
The foreground extinction is not uniform and strong
(around 2--3~mag in the $K$ band, $2.2~\mu{\mathrm m}$).
Thus, infrared observations are required
in order to study stars in the GC region.
In fact long-term infrared observations have made it possible to 
monitor stellar motions around the central black hole
(\citealt{Ghez-2008}; \citealt{Gillessen-2009}, and references therein). 
These and other data were used to search for variable stars 
in the few parsec (or smaller) region around Sgr~A$^*$
(\citealt{Tamura-1996};
Ott, Eckart \& Genzel, 1999; Peeples, Stanek \& DePoy, 2007;
\citealt{Rafelski-2007}).
However, no Cepheids were found in these works.

We carried out near-IR observations 
to investigate stellar variability in the GC region.
Our survey covered a much wider area, $20^\prime \times 30^\prime$,
than the previous monitoring observations.
A large number of long-period variables including Miras
were found in the survey region
(\citeauthor{Matsunaga-2009b}~\citeyear{Matsunaga-2009b}, =Paper~I),
and we discovered three CCEPs, the first of this type in the GC region
(\citeauthor{Matsunaga-2011b}~\citeyear{Matsunaga-2011b}, =Paper~II).
In the present paper we describe our data analysis and
a catalogue of the short period variables in the field,
and also discuss their nature as well as the distance to the GC.

\section{Observation and data reduction}

Observations were conducted using the IRSF 1.4~m telescope and
the SIRIUS camera (\citealt{Nagashima-1999}; \citealt{Nagayama-2003})
which collects images in 
the $J~(1.25~\mu\mathrm{m})$, $H~(1.63~\mu\mathrm{m})$
and $\Ks~(2.14~\mu\mathrm{m})$ bands, simultaneously.
The observed field composed of 12 fields-of-view of IRSF/SIRIUS
covered $20^\prime \times 30^\prime$ around the GC (Table~\ref{tab:Fields}).
Observations at about 90 epochs were made between 2001 and 2008 
of which the majority were obtained in 2005 and 2006.
We used this dataset in Paper~I and II,
and further observational details are found there.

\begin{table}
\begin{minipage}{80mm}
\caption{The observed fields. The central coordinates,
the numbers of monitorings 
and the numbers of short-period variables are listed.
\label{tab:Fields}}
\begin{center}
\begin{tabular}{ccccc}
\hline
Field   &  RA~(J2000) & Dec.~(J2000.0) & $N_{\mathrm{obs}}$ & $N_{\mathrm{Short}}$ \\
\hline
1745$-$2900A & 17:46:10.5 & $-$28:53:47.8 & 94 & 3 \\
1745$-$2900B & 17:45:40.0 & $-$28:53:47.8 & 92 & 6 \\
1745$-$2900C & 17:45:09.5 & $-$28:53:47.8 & 90 & 3 \\
1745$-$2900D & 17:46:10.5 & $-$29:00:28.0 & 93 & 3 \\
1745$-$2900E & 17:45:40.0 & $-$29:00:28.0 & 91 & 6$^\dagger$ \\
1745$-$2900F & 17:45:09.5 & $-$29:00:28.0 & 89 & 5$^\dagger$ \\
1745$-$2900G & 17:46:10.5 & $-$29:07:07.8 & 87 & 4 \\
1745$-$2900H & 17:45:40.0 & $-$29:07:07.8 & 89 & 3 \\
1745$-$2900I & 17:45:09.5 & $-$29:07:07.8 & 83 & 4 \\
1745$-$2840G & 17:46:10.5 & $-$28:47:07.9 & 85 & 6 \\
1745$-$2840H & 17:45:40.0 & $-$28:47:07.9 & 85 & 1 \\
1745$-$2840I & 17:45:09.5 & $-$28:47:07.9 & 60 & 2 \\
Total      & & & & 45$^\ddagger$ \\
\hline
\end{tabular}
\end{center}
$^\dagger$~One object, \#15 in Table~\ref{tab:catalogue},
was detected in the overlapping region of both fields.
$^\ddagger$~We do not include the duplicate detection
in the neighbouring fields.
\end{minipage}
\end{table}

The basic data analysis was done in the same manner as in Paper~I.
In short, point-spread-function (PSF) fitting photometry was performed
on every images using the {\small IRAF/DAOPHOT} package,
and variable stars were searched for by combining the
time-series sets of $\JHK$ magnitudes. The standard deviations (SDs)
were calculated for repeated measurements of individual stars.
We then looked for variable stars with SD
more than three times larger than the median value of SDs
in the corresponding magnitude range.
The variability search was done using the three band datasets independently,
so that we could find variables even if they are visible only in
one of the $\JHK$ bands.

The saturation limits are 9.5, 9.5 and 9.0~mag, and the detection limits
are around 16.4, 14.5 and 13.1~mag in $J$, $H$ and $\Ks$, respectively.
The definition of these values is described in Paper~I, but
the detection limits vary across the survey region depending
on the crowdedness. Especially, the central region around Sgr~A$^*$
is so crowded that the accuracy of our photometric measurements,
with the typical seeing of $\sim 1^{\prime\prime}$, is rather limited.
The detection limit also changes from frame to frame depending on
the weather condition. Therefore, the above limiting magnitudes
should be considered only as typical values.

While we catalogued 1364 long-period variables in Paper~I, 
variables with period shorter than 60~days
are presented in this paper.
Periods $P$ were determined 
by fitting the following fourth-order Fourier series,
\begin{equation}
m(t) = A_0 + \sum_{i=1}^4  A_i \cos ( 2\pi it / P+\phi_i ).
\label{eq:fourier}
\end{equation}
The light curves of Cepheids, which are our main targets,
are known to be fitted by such Fourier series well (e.g.~\citealt{Laney-1993}).

\section{Results}

\subsection{Detection of short-period variables}
\label{sec:catalogue}

\begin{table*}
\begin{minipage}{170mm}
\caption{
The catalogue of short-period variables detected towards the Galactic Centre.
After the numbering ID between \#1 and \#45, the ID combining
RA and Dec.~(J2000.0) follows. Then listed are galactic coordinates,
$\JHK$ mean magnitudes, peak-to-valley
amplitudes, $M$flag (see the text) and periods.
The classifications are also indicated as: Cep(I)$=$classical Cepheid,
Cep(II)$=$type II Cepheid, Ecl$=$eclipsing binary
(see also Table~\ref{tab:classification}). \#3 may be either of an RR~Lyr
or a $\delta$~Sct star. \#34 seems to be a kind of Cepheid, possibly
an anomalous Cepheid, but it is unclear which group it belongs to.
For eclipsing binaries except \#30, half the orbital periods are listed.
\label{tab:catalogue}}
\begin{center}
\begin{tabular}{rcrrrrrrrrccc}
\hline
\multicolumn{1}{c}{No.} & ID & \multicolumn{1}{c}{$l$} & \multicolumn{1}{c}{$b$} & \multicolumn{1}{c}{$J$} & \multicolumn{1}{c}{$H$} & \multicolumn{1}{c}{$\Ks$} & \multicolumn{1}{c}{$\Delta J$} & \multicolumn{1}{c}{$\Delta H$} & \multicolumn{1}{c}{$\Delta \Ks$} & Mflag & Period & Type \\
 & & \multicolumn{1}{c}{($\circ$)} & \multicolumn{1}{c}{($\circ$)} & \multicolumn{1}{c}{(mag)} & \multicolumn{1}{c}{(mag)} & \multicolumn{1}{c}{(mag)} & \multicolumn{1}{c}{(mag)} & \multicolumn{1}{c}{(mag)} & \multicolumn{1}{c}{(mag)} &  & (d) &  \\
\hline
 1 & 17445710$-$2910057 & $-0.2742$ & $+0.0036$ & 15.55 & 14.35 &  \multicolumn{1}{c}{---}  & 0.56 & 0.52 &  \multicolumn{1}{c}{---}  & 003 & 0.3613   & Ecl  \\
 2 & 17445906$-$2851235 & $-0.0046$ & $+0.1602$ & 15.57 & 14.08 & 12.73 & 1.22 & 1.03 & 0.62 & 777 & 7.46     & Cep(II)  \\
 3 & 17445910$-$2909440 & $-0.2652$ & $+0.0005$ & 13.76 & 12.53 & 11.72 & 0.23 & 0.13 & 0.31 & 000 & 0.265    & RR/DS  \\
 4 & 17450132$-$2848213 & $+0.0428$ & $+0.1796$ &  \multicolumn{1}{c}{---}  & 14.46 & 12.87 &  \multicolumn{1}{c}{---}  & 0.71 & 0.69 & 300 & 12.544   & Ecl  \\
 5 & 17450204$-$2857215 & $-0.0838$ & $+0.0990$ & 14.58 & 14.06 &  \multicolumn{1}{c}{---}  & 0.66 & 0.56 &  \multicolumn{1}{c}{---}  & 003 & 0.17733  & Ecl  \\
 6 & 17450754$-$2906573 & $-0.2097$ & $-0.0015$ &  \multicolumn{1}{c}{---}  & 14.13 & 12.51 &  \multicolumn{1}{c}{---}  & 0.61 & 0.68 & 300 & 15.097   & Cep(II)  \\
 7 & 17450913$-$2859417 & $-0.1035$ & $+0.0567$ & 16.37 & 13.02 & 11.33 & 0.68 & 0.47 & 0.40 & 000 & 52.224   & Cep(II)  \\
 8 & 17451032$-$2904526 & $-0.1749$ & $+0.0079$ & 14.08 & 13.39 & 12.42 & 0.35 & 0.44 & 0.35 & 077 & 0.21968  & Ecl  \\
 9 & 17451383$-$2844443 & $+0.1181$ & $+0.1721$ &  \multicolumn{1}{c}{---}  & 15.19 & 13.69 &  \multicolumn{1}{c}{---}  & 0.53 & 0.50 & 300 & 4.747    & Cep(II)  \\
10 & 17451719$-$2857531 & $-0.0624$ & $+0.0474$ &  \multicolumn{1}{c}{---}  & 14.49 & 12.47 &  \multicolumn{1}{c}{---}  & 0.74 & 0.86 & 300 & 24.09    & Cep(II)  \\
11 & 17451764$-$2851372 & $+0.0275$ & $+0.1004$ &  \multicolumn{1}{c}{---}  & 14.91 & 13.30 &  \multicolumn{1}{c}{---}  & 0.34 & 0.35 & 300 & 8.2713   & Cep(II)  \\
12 & 17452092$-$2858186 & $-0.0614$ & $+0.0321$ & 14.30 & 13.53 &  \multicolumn{1}{c}{---}  & 0.87 & 0.77 &  \multicolumn{1}{c}{---}  & 003 & 0.15869  & Ecl  \\
13 & 17452219$-$2853583 & $+0.0027$ & $+0.0658$ & 12.58 & 12.29 & 12.11 & 0.34 & 0.34 & 0.33 & 000 & 1.6094   & Ecl  \\
14 & 17452573$-$2909397 & $-0.2137$ & $-0.0815$ &  \multicolumn{1}{c}{---}  & 14.37 & 12.76 &  \multicolumn{1}{c}{---}  & 0.43 & 0.41 & 377 & 1.0984   & Ecl  \\
15 & 17452600$-$2900037 & $-0.0766$ & $+0.0010$ & 15.69 & 12.93 & 11.36 & 0.90 & 0.92 & 0.96 & 000 & 50.46    & Cep(II)  \\
16 & 17452837$-$2858221 & $-0.0480$ & $+0.0084$ & 15.02 & 13.94 &  \multicolumn{1}{c}{---}  & 0.54 & 0.52 &  \multicolumn{1}{c}{---}  & 003 & 1.5838   & Ecl  \\
17 & 17452987$-$2854290 & $+0.0101$ & $+0.0375$ & 14.97 & 12.18 & 10.67 & 0.29 & 0.26 & 0.27 & 000 & 1.6448   & Ecl  \\
18 & 17453089$-$2903105 & $-0.1116$ & $-0.0412$ & 16.36 & 12.44 & 10.35 & 0.68 & 0.44 & 0.51 & 000 & 22.76    & Cep(I)  \\
19 & 17453148$-$2859531 & $-0.0637$ & $-0.0145$ & 10.98 & 10.84 & 10.70 & 0.10 & 0.10 & 0.18 & 000 & 3.6301   & Cep(II)  \\
20 & 17453227$-$2902552 & $-0.1054$ & $-0.0433$ & 15.42 & 12.00 & 10.17 & 0.60 & 0.46 & 0.57 & 000 & 19.96    & Cep(I)  \\
21 & 17454075$-$2852367 & $+0.0574$ & $+0.0198$ &  \multicolumn{1}{c}{---}  & 14.93 & 13.31 &  \multicolumn{1}{c}{---}  & 0.39 & 0.47 & 300 & 0.55648  & Ecl  \\
22 & 17454904$-$2856450 & $+0.0142$ & $-0.0419$ & 13.81 & 13.40 & 13.02 & 0.55 & 0.62 & 0.71 & 007 & 0.41278  & Ecl  \\
23 & 17455015$-$2855069 & $+0.0396$ & $-0.0312$ &  \multicolumn{1}{c}{---}  & 14.25 & 12.53 &  \multicolumn{1}{c}{---}  & 0.36 & 0.42 & 377 & 1.628    & Ecl  \\
24 & 17455150$-$2903392 & $-0.0793$ & $-0.1094$ & 14.18 & 13.72 & 13.58 & 0.40 & 0.42 & 0.54 & 000 & 0.24946  & Ecl  \\
25 & 17455257$-$2900004 & $-0.0254$ & $-0.0811$ & 17.05 & 14.03 & 12.19 & 0.89 & 0.57 & 0.58 & 700 & 1.7092   & Ecl  \\
26 & 17455318$-$2856206 & $+0.0279$ & $-0.0512$ &  \multicolumn{1}{c}{---}  & 14.28 & 12.40 &  \multicolumn{1}{c}{---}  & 0.84 & 0.85 & 000 & 16.1     & Cep(II)  \\
27 & 17455325$-$2904069 & $-0.0826$ & $-0.1189$ &  \multicolumn{1}{c}{---}  & 14.53 & 12.89 &  \multicolumn{1}{c}{---}  & 0.41 & 0.42 & 300 & 1.7316   & Ecl  \\
28 & 17455413$-$2845032 & $+0.1904$ & $+0.0437$ &  \multicolumn{1}{c}{---}  & 14.62 & 12.95 &  \multicolumn{1}{c}{---}  & 0.77 & 0.73 & 300 & 15.543   & Cep(II)  \\
29 & 17455482$-$2854382 & $+0.0553$ & $-0.0415$ &  \multicolumn{1}{c}{---}  & 15.29 & 13.58 &  \multicolumn{1}{c}{---}  & 0.45 & 0.61 & 377 & 10.26    & Cep(II)  \\
30 & 17460164$-$2855155 & $+0.0594$ & $-0.0682$ & 13.40 & 10.64 &  9.16 & 0.40 & 0.42 & 0.32 & 000 & 26.792   & Ecl  \\
31 & 17460200$-$2852506 & $+0.0944$ & $-0.0484$ &  \multicolumn{1}{c}{---}  & 12.98 & 11.37 &  \multicolumn{1}{c}{---}  & 0.61 & 0.68 & 377 & 40.13    & Cep(II)  \\
32 & 17460601$-$2846551 & $+0.1864$ & $-0.0095$ & 15.63 & 12.04 & 10.18 & 0.58 & 0.45 & 0.44 & 000 & 23.538   & Cep(I)  \\
33 & 17460637$-$2909442 & $-0.1377$ & $-0.2084$ & 12.68 & 10.91 & 10.07 & 0.12 & 0.14 & 0.10 & 000 & 18.96    & Cep(II)  \\
34 & 17461000$-$2855325 & $+0.0712$ & $-0.0967$ & 15.01 & 12.28 & 10.79 & 0.21 & 0.17 & 0.19 & 077 & 2.1932   & Cep(?)  \\
35 & 17461007$-$2905173 & $-0.0674$ & $-0.1814$ & 16.10 &  \multicolumn{1}{c}{---}  &  \multicolumn{1}{c}{---}  & 0.73 &  \multicolumn{1}{c}{---}  &  \multicolumn{1}{c}{---}  & 033 & 0.14612  & Ecl  \\
36 & 17461044$-$2903183 & $-0.0385$ & $-0.1653$ & 12.43 & 11.89 & 11.18 & 0.79 & 0.72 & 0.56 & 077 & 0.97209  & Ecl  \\
37 & 17461171$-$2850001 & $+0.1533$ & $-0.0540$ & 16.08 & 13.45 & 11.91 & 0.74 & 0.56 & 0.60 & 000 & 0.94255  & Ecl  \\
38 & 17461252$-$2848526 & $+0.1709$ & $-0.0468$ &  \multicolumn{1}{c}{---}  & 14.24 & 12.24 &  \multicolumn{1}{c}{---}  & 0.59 & 0.54 & 300 & 1.6486   & Ecl  \\
39 & 17461356$-$2848351 & $+0.1770$ & $-0.0475$ &  \multicolumn{1}{c}{---}  &  \multicolumn{1}{c}{---}  & 12.75 &  \multicolumn{1}{c}{---}  &  \multicolumn{1}{c}{---}  & 0.86 & 337 & 31.17    & Cep(II)  \\
40 & 17461357$-$2859023 & $+0.0282$ & $-0.1381$ &  \multicolumn{1}{c}{---}  & 14.91 & 13.03 &  \multicolumn{1}{c}{---}  & 1.00 & 0.84 & 300 & 19.014   & Cep(II)  \\
41 & 17461447$-$2849002 & $+0.1728$ & $-0.0539$ & 13.80 & 12.57 & 10.98 & 0.73 & 0.53 & 0.27 & 077 & 0.14161  & Ecl  \\
42 & 17461626$-$2850125 & $+0.1590$ & $-0.0700$ & 15.63 & 12.97 & 11.39 & 0.77 & 0.70 & 0.72 & 000 & 1.66284  & Ecl  \\
43 & 17462426$-$2908288 & $-0.0860$ & $-0.2531$ &  \multicolumn{1}{c}{---}  & 14.06 & 12.49 &  \multicolumn{1}{c}{---}  & 0.60 & 0.69 & 300 & 13.52    & Ecl  \\
44 & 17462642$-$2857079 & $+0.0797$ & $-0.1616$ & 14.33 & 13.53 & 13.23 & 0.30 & 0.35 & 0.37 & 000 & 0.41546  & Ecl  \\
45 & 17462846$-$2908562 & $-0.0846$ & $-0.2702$ & 17.17 & 14.03 & 12.46 & 1.34 & 0.99 & 1.10 & 000 & 24.406   & Cep(II)  \\
\hline
\end{tabular}
\end{center}
\end{minipage}
\end{table*}

\begin{figure*}
\begin{minipage}{160mm}
\begin{center}
\includegraphics[clip,width=0.99\hsize]{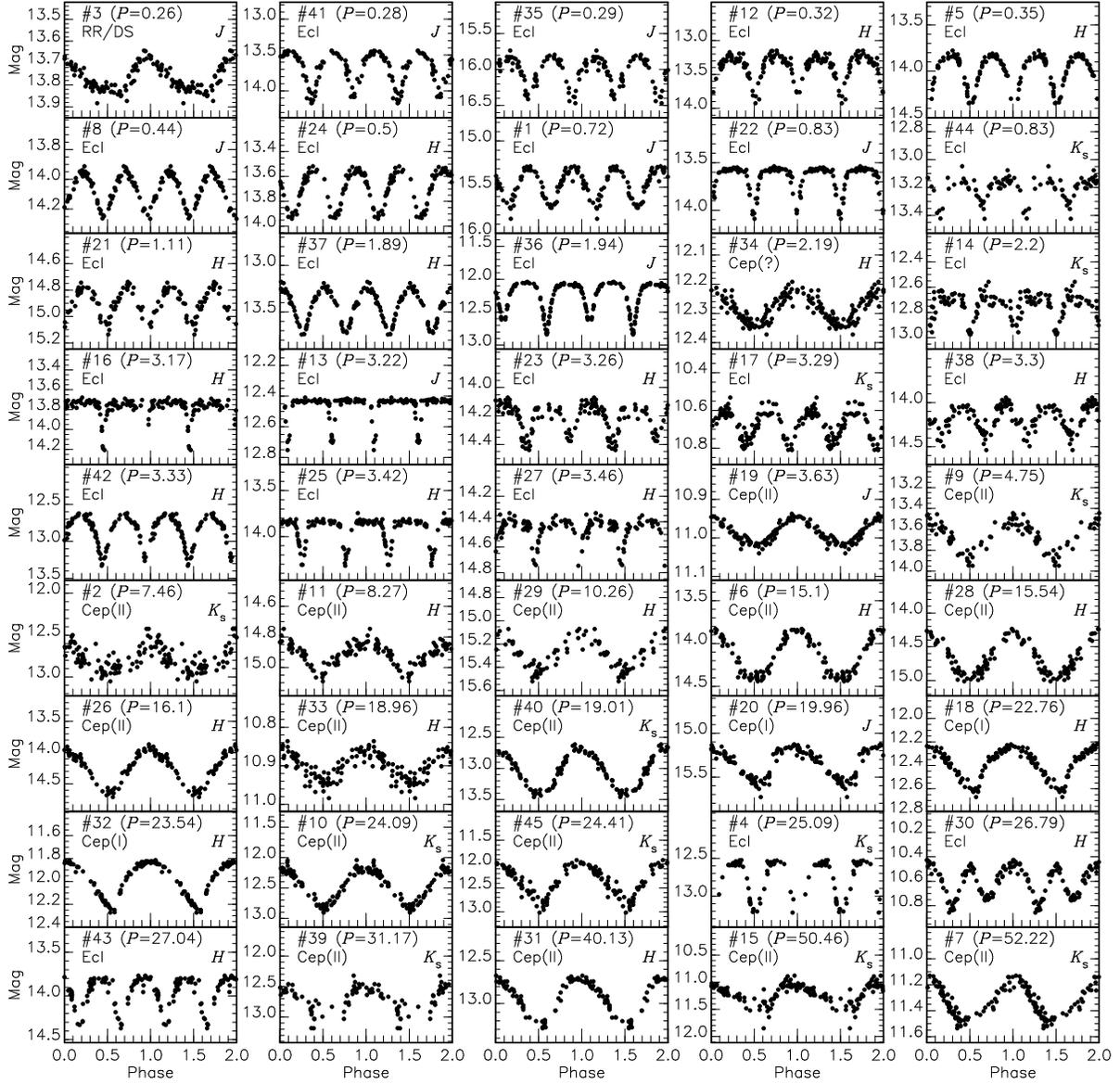}
\end{center}
\caption{
Light curves of the catalogued variables plotted in the ascending order of period.
For eclipsing variables, the orbital periods (twice the periods listed in Table~\ref{tab:catalogue}) are used.
The ID, the variable type and the (rounded) period are indicated at the top
of each panel
as well as the name of filter ($J$, $H$ or $\Ks$) used in the plot.
\label{fig:LCs}}
\end{minipage}
\end{figure*}

We detected 45 variable stars with period between 0.14 and 52.1~d. 
The number of the objects found in each field-of-view is indicated
in Table~\ref{tab:Fields}.
Table~\ref{tab:catalogue} lists their IDs, galactic coordinates,
mean magnitudes, amplitudes and periods. The mean magnitudes are 
intensity-scale means of maximum
and minimum, and the amplitudes refer to peak-to-valley variations.
The $\JHK$ time-series data obtained for all the catalogued variables
are compiled in one text file and each line includes the ID number,
the modified Julian date (MJD) and the $\JHK$ for each measurement.
Table~\ref{tab:LCdata} shows the first 10 lines as a sample
of the full version to be published online.
Fig.~\ref{fig:LCs} plots their folded light curves in
the ascending order of period.
Because the light curves of eclipsing variables are often
nearly symmetrical, a fit of the Fourier series (eq.~\ref{eq:fourier})
tends to yield half the orbital period and this is listed
in Table~\ref{tab:catalogue} except in the case of \#30
whose light curve is significantly asymmetric.
The orbital periods are used in Fig.~\ref{fig:LCs}.

We did not always detect the variables in all of the $\JHK$ bands.
Table~\ref{tab:catalogue} includes $M$flag which we also used
in Paper~I to show the reasons of non-detection or the qualities
of the listed magnitudes.
In this work, only the flag numbers 0, 3 and 7 are relevant.
The flags 0 and 3 respectively indicate that a mean magnitude
was obtained properly and that some measurements were affected
by the detection limit leading to an uncertain mean magnitude.
The flag 7 is newly defined to indicate that 
the photometry of the object is affected by the crowding.
None of our objects was too bright, and none was located too close
to the edge of the detector.
The light curves in Fig.~\ref{fig:LCs} indicates that
the entire variations from minima to maxima were sampled well enough
to estimate mean magnitudes except for the faintest cases.

As we see in Fig.~\ref{fig:LCs}, our sample includes different types of variables.
In order to determine the variable types,
shapes of the light curves are discussed in Section~\ref{sec:LCshape}.
For CCEPs and T2Cs, as briefly discussed in Paper II,
we also consider their absolute magnitudes and 
the expected distances (Section~\ref{sec:CepDistance}).
In Section~\ref{sec:Classification}, we summarise the classification and
compare some features among variable types.

\begin{table}
\begin{minipage}{80mm}
\caption{
Light variations for the catalogued variables.
When the magnitudes are not available we put 99.99 instead.
This is the first 10 lines of the full catalogue (3707~lines),
which will be available in the online version of the article 
(see Supporting Information).
\label{tab:LCdata}}
\begin{center}
\begin{tabular}{ccrrr}
\hline
No. & MJD & \multicolumn{1}{c}{$J$} & \multicolumn{1}{c}{$H$} & \multicolumn{1}{c}{$\Ks$} \\
\hline
 1 & 52343.1514 & 15.60 & 14.49 & 99.99 \\
 1 & 53482.0694 & 15.28 & 14.22 & 99.99 \\
 1 & 53537.1167 & 15.57 & 99.99 & 99.99 \\
 1 & 53540.8287 & 15.69 & 14.56 & 99.99 \\
 1 & 53545.8959 & 15.67 & 14.49 & 99.99 \\
 1 & 53545.9758 & 15.32 & 14.24 & 99.99 \\
 1 & 53548.8325 & 15.48 & 99.99 & 99.99 \\
 1 & 53548.9662 & 15.32 & 14.20 & 99.99 \\
 1 & 53549.8964 & 15.53 & 14.39 & 99.99 \\
 1 & 53550.0148 & 15.32 & 14.17 & 99.99 \\
\hline
\end{tabular}
\end{center}
\end{minipage}
\end{table}

\subsection{Shapes of the light curves}
\label{sec:LCshape}

In order to give a quantitative description of the light curve shape,
the parameters,
\begin{eqnarray}
R_{21} = A_2 / A_1, \label{eq:R21} \\
\phi_{21} = \phi_2 - 2 \phi_1, \label{eq:phi21} \\
R_{31} = A_3 / A_1, \label{eq:R31} \\
\phi_{31} = \phi_3 - 3 \phi_1, \label{eq:phi31}
\label{eq:Four31}
\end{eqnarray}
are considered for each light curve based on the fitted Fourier series
(eq.~\ref{eq:fourier}).
These parameters are listed in Table~\ref{tab:Fourier}.

\begin{table*}
\begin{minipage}{150mm}
\caption{
The list of Fourier parameters for the observed light curves. Four parameters ($R_{21}, \phi_{21}, R_{31}$ and $\phi_{31}$, see eq.~\ref{eq:R21}--\ref{eq:phi31} for the definition) are listed for each of the $\JHK$ bands.
The $\phi$ values are given in the unit of $\pi$.
\label{tab:Fourier}}
\begin{center}
\begin{tabular}{rcrrrrrrrrrrrrr}
\hline
\multicolumn{1}{c}{No.} & \multicolumn{1}{c}{$\log P$} & \multicolumn{4}{c}{$J$-band} & \multicolumn{4}{c}{$H$-band} & \multicolumn{4}{c}{$\Ks$-band} \\
& & $R_{21}$ & $\phi_{21}$ & $R_{31}$ & $\phi_{31}$ & $R_{21}$ & $\phi_{21}$ & $R_{31}$ & $\phi_{31}$ & $R_{21}$ & $\phi_{21}$ & $R_{31}$ & $\phi_{31}$ \\
\hline
     1 & $-0.442^\dagger$ & 0.176 & 1.983 & 0.030 & 1.550 & 0.174 & 2.004 & 0.075 & 0.064 &  ---  &  ---  &  ---  &  ---  \\
     2 & $ 0.873$ & 0.246 & 2.923 & 0.157 & 0.212 & 0.336 & 2.834 & 0.120 & 0.204 & 0.364 & 2.907 & 0.192 & 0.021 \\
     3 & $-0.577$ & 0.312 & 1.181 & 0.181 & 0.690 & 0.309 & 1.304 & 0.242 & 0.862 & 0.526 & 1.484 & 0.144 & 1.518 \\
     4 & $ 1.098^\dagger$ &  ---  &  ---  &  ---  &  ---  & 0.301 & 2.055 & 0.072 & 0.739 & 0.268 & 2.067 & 0.074 & 0.728 \\
     5 & $-0.751^\dagger$ & 8.190 & 1.913 & 1.076 & 1.961 & 3.818 & 1.918 & 0.715 & 1.951 &  ---  &  ---  &  ---  &  ---  \\
     6 & $ 1.179$ &  ---  &  ---  &  ---  &  ---  & 0.038 & 1.403 & 0.051 & 0.774 & 0.044 & 1.002 & 0.113 & 0.812 \\
     7 & $ 1.718$ & 0.283 & 2.615 & 0.096 & 0.098 & 0.184 & 2.696 & 0.036 & 1.956 & 0.147 & 2.665 & 0.005 & 1.757 \\
     8 & $-0.658^\dagger$ & 0.087 & 2.016 & 0.064 & 0.140 & 0.132 & 1.936 & 0.091 & 0.848 & 0.384 & 1.998 & 0.247 & 1.576 \\
     9 & $ 0.676$ &  ---  &  ---  &  ---  &  ---  & 0.259 & 1.986 & 0.075 & 0.184 & 0.116 & 2.101 & 0.051 & 0.189 \\
    10 & $ 1.382$ &  ---  &  ---  &  ---  &  ---  & 0.160 & 1.869 & 0.029 & 0.038 & 0.101 & 1.822 & 0.029 & 1.258 \\
    11 & $ 0.918$ &  ---  &  ---  &  ---  &  ---  & 0.227 & 2.188 & 0.088 & 0.245 & 0.183 & 2.378 & 0.062 & 0.046 \\
    12 & $-0.799^\dagger$ & 5.306 & 2.023 & 0.526 & 1.955 & 4.916 & 1.983 & 0.615 & 1.894 &  ---  &  ---  &  ---  &  ---  \\
    13 & $ 0.207^\dagger$ & 0.683 & 1.912 & 0.568 & 1.877 & 0.733 & 1.923 & 0.585 & 1.888 & 0.689 & 1.949 & 0.621 & 1.932 \\
    14 & $ 0.041^\dagger$ &  ---  &  ---  &  ---  &  ---  & 0.493 & 2.043 & 0.308 & 1.833 & 0.702 & 1.908 & 0.237 & 1.707 \\
    15 & $ 1.703$ & 0.176 & 1.585 & 0.064 & 0.655 & 0.359 & 1.595 & 0.088 & 0.934 & 0.316 & 1.536 & 0.105 & 0.829 \\
    16 & $ 0.200^\dagger$ & 0.946 & 2.063 & 0.934 & 0.072 & 0.856 & 2.036 & 0.819 & 0.043 &  ---  &  ---  &  ---  &  ---  \\
    17 & $ 0.216^\dagger$ & 0.332 & 2.013 & 0.077 & 1.987 & 0.232 & 1.973 & 0.058 & 1.737 & 0.240 & 2.094 & 0.068 & 0.830 \\
    18 & $ 1.357$ & 0.236 & 1.729 & 0.240 & 1.059 & 0.189 & 1.896 & 0.138 & 1.740 & 0.239 & 1.922 & 0.182 & 1.710 \\
    19 & $ 0.560$ & 0.118 & 1.337 & 0.068 & 1.331 & 0.137 & 1.243 & 0.010 & 0.926 & 0.541 & 1.030 & 0.087 & 1.152 \\
    20 & $ 1.300$ & 0.311 & 1.679 & 0.135 & 1.260 & 0.200 & 1.849 & 0.107 & 1.500 & 0.349 & 1.753 & 0.066 & 1.599 \\
    21 & $-0.255^\dagger$ &  ---  &  ---  &  ---  &  ---  & 0.127 & 2.101 & 0.090 & 0.048 & 0.233 & 1.865 & 0.129 & 1.737 \\
    22 & $-0.384^\dagger$ & 0.692 & 2.030 & 0.423 & 0.069 & 0.696 & 1.979 & 0.430 & 0.006 & 0.497 & 2.117 & 0.854 & 0.329 \\
    23 & $ 0.212^\dagger$ &  ---  &  ---  &  ---  &  ---  & 1.853 & 2.542 & 0.189 & 0.838 & 1.713 & 2.694 & 0.114 & 1.046 \\
    24 & $-0.603^\dagger$ & 0.161 & 2.003 & 0.025 & 1.545 & 0.189 & 2.013 & 0.015 & 0.797 & 0.253 & 2.023 & 0.051 & 0.319 \\
    25 & $ 0.233^\dagger$ & 0.661 & 2.035 & 0.310 & 0.169 & 0.655 & 2.029 & 0.370 & 0.114 & 0.619 & 2.004 & 0.328 & 0.035 \\
    26 & $ 1.207$ &  ---  &  ---  &  ---  &  ---  & 0.098 & 1.798 & 0.049 & 0.332 & 0.041 & 1.301 & 0.035 & 0.810 \\
    27 & $ 0.238^\dagger$ &  ---  &  ---  &  ---  &  ---  & 0.586 & 2.064 & 0.513 & 0.210 & 0.805 & 2.118 & 0.620 & 0.233 \\
    28 & $ 1.192$ &  ---  &  ---  &  ---  &  ---  & 0.015 & 1.944 & 0.056 & 0.673 & 0.067 & 1.129 & 0.040 & 0.874 \\
    29 & $ 1.011$ &  ---  &  ---  &  ---  &  ---  & 0.108 & 2.278 & 0.041 & 0.511 & 0.093 & 1.753 & 0.185 & 0.738 \\
    30 & $ 1.428$ & 1.969 & 1.022 & 0.856 & 1.179 & 1.818 & 1.016 & 0.734 & 1.160 & 1.656 & 2.980 & 0.566 & 1.157 \\
    31 & $ 1.603$ &  ---  &  ---  &  ---  &  ---  & 0.151 & 1.749 & 0.049 & 0.952 & 0.118 & 1.719 & 0.084 & 0.843 \\
    32 & $ 1.372$ & 0.299 & 1.671 & 0.189 & 1.422 & 0.253 & 1.879 & 0.129 & 1.698 & 0.210 & 1.910 & 0.101 & 1.735 \\
    33 & $ 1.278$ & 0.036 & 2.932 & 0.055 & 1.459 & 0.034 & 1.853 & 0.121 & 1.310 & 0.085 & 1.345 & 0.109 & 1.236 \\
    34 & $ 0.341$ & 0.124 & 1.771 & 0.180 & 1.735 & 0.086 & 1.757 & 0.029 & 0.409 & 0.049 & 1.722 & 0.068 & 1.395 \\
    35 & $-0.835^\dagger$ & 0.318 & 2.076 & 0.148 & 0.093 &  ---  &  ---  &  ---  &  ---  &  ---  &  ---  &  ---  &  ---  \\
    36 & $-0.012^\dagger$ & 0.458 & 1.978 & 0.265 & 1.985 & 0.461 & 1.985 & 0.252 & 0.003 & 0.414 & 1.935 & 0.234 & 1.949 \\
    37 & $-0.026^\dagger$ & 0.155 & 2.014 & 0.100 & 0.186 & 0.166 & 1.991 & 0.094 & 1.960 & 0.128 & 1.902 & 0.135 & 1.912 \\
    38 & $ 0.217^\dagger$ &  ---  &  ---  &  ---  &  ---  & 0.113 & 1.789 & 0.121 & 1.735 & 0.114 & 1.684 & 0.075 & 1.489 \\
    39 & $ 1.494$ &  ---  &  ---  &  ---  &  ---  &  ---  &  ---  &  ---  &  ---  & 0.099 & 1.414 & 0.194 & 0.897 \\
    40 & $ 1.279$ &  ---  &  ---  &  ---  &  ---  & 0.142 & 1.902 & 0.064 & 0.275 & 0.087 & 1.662 & 0.051 & 0.793 \\
    41 & $-0.849^\dagger$ & 0.328 & 1.962 & 0.112 & 0.039 & 0.357 & 1.877 & 0.108 & 1.829 & 0.324 & 1.747 & 0.058 & 1.061 \\
    42 & $ 0.221^\dagger$ & 0.324 & 1.999 & 0.152 & 1.978 & 0.362 & 1.970 & 0.175 & 0.008 & 0.348 & 1.946 & 0.158 & 0.002 \\
    43 & $ 1.131^\dagger$ &  ---  &  ---  &  ---  &  ---  & 0.382 & 1.966 & 0.110 & 1.954 & 0.318 & 1.955 & 0.068 & 1.866 \\
    44 & $-0.381^\dagger$ & 0.585 & 1.922 & 0.255 & 1.888 & 0.779 & 2.031 & 0.327 & 0.072 & 0.548 & 2.022 & 0.340 & 1.827 \\
    45 & $ 1.387$ & 0.197 & 1.920 & 0.122 & 1.312 & 0.157 & 1.785 & 0.093 & 1.502 & 0.171 & 1.752 & 0.108 & 1.618 \\
\hline
\end{tabular}
\end{center}
$\dagger$~Half the orbital periods are given for all the eclipsing binaries
except \#30.
\end{minipage}
\end{table*}

\begin{figure*}
\begin{minipage}{170mm}
\begin{center}
\includegraphics[clip,width=0.98\hsize]{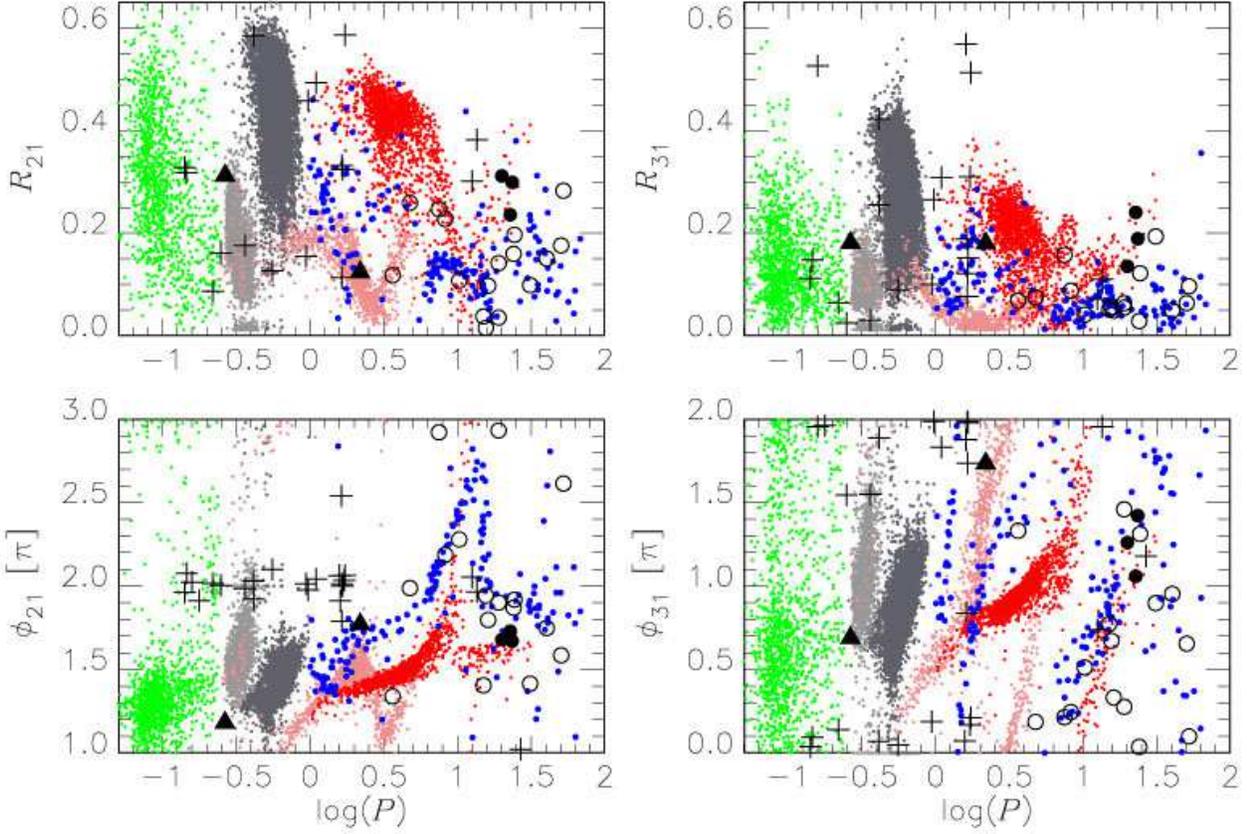}
\end{center}
\caption{
Fourier parameters (eq.~\ref{eq:R21}--\ref{eq:phi31}) plotted against period.
For eclipsing binaries except \#30, half of the orbital periods are used.
Black large symbols indicate our objects:
filled circles for CCEPs, open circles for T2Cs,
'$+$' symbols for binaries and triangles for others.
The small dots indicate the OGLE-III variable stars of different types
with different colours: red for CCEPs pulsating in the fundamental
mode, pink for CCEPs in the overtone mode, blue for T2Cs in the fundamental
mode, dark grey for RR~Lyr stars in the fundamental mode, light grey
for those in the overtone mode, and green for $\delta$~Sct stars.
The Fourier parameters were obtained with the $I$-band light curves
for the OGLE-III stars, while the light curve in the shortest wavelength
available was used for each of our objects.
\label{fig:Fours}}
\end{minipage}
\end{figure*}

We also consider the above Fourier parameters for the variables in
the Large Magellanic Cloud (LMC), found in the Optical Gravitational
Lensing Experiment (OGLE-III), to compare with our objects.
In Fig.~\ref{fig:Fours}, different types of the LMC variables
are plotted in different colours:
CCEP \citep{Soszynski-2008a}, T2Cs \citep{Soszynski-2008b},
RR Lyr stars \citep{Soszynski-2009},
and $\delta$~Sct stars \citep{Poleski-2010}.
Among the $\delta$~Sct reported by \citet{Poleski-2010},
single-mode stars without the uncertainty flag are used.
Because they included only $R_{21}$ and $\phi_{21}$, we calculated
$R_{31}$ and $\phi_{31}$ using their photometric data.

The OGLE-III light curves were taken in the $I$-band.
The available light curves are rather limited in $\JHK$, but the result in
\citet{Laney-1993} suggests that, at least, $J$-band light curves are
similar to the $I$-band ones for CCEPs.
We plot the parameters for $J$-band light curves
whenever possible for our objects. Those for $H$-band are used in other cases,
but for the object with neither of $J$ and $H$ light curves
the $\Ks$-band parameters are considered. 
The second column of Table~\ref{tab:classification} indicates
the variability types judged by the light curve shapes.

\begin{table}
\begin{minipage}{80mm}
\caption{
Classification of variable stars.
The second column indicates the variable types whose light curve shapes
are consistent with those of the objects. 
As the variable types, I and II stand for CCEPs and T2Cs, respectively,
while Ecl stands for eclipsing binaries. 
The $\AK$ and $D$ indicate approximate extinctions and distances
for Cepheids (I$=$CCEPs, II$=$T2Cs).
In the last column the concluded variable types are listed.
See the details in the text where some comments on individual objects
are also given.
\label{tab:classification}}
\begin{center}
\begin{tabular}{lcccc}
\hline
No. & LC    & $\AK$ & $D$   & Type \\
    & shape & (mag) & (kpc) &      \\
\hline
 1 & Ecl    & ---   & ---                 & Ecl     \\
 2 & I/II:  & 1.3   & 23.9 (I),  9.0 (II) & Cep(II) \\
 3 & RR/DS  & ---   & ---                 & RR/DS  $\dagger$  \\
 4 & Ecl     & ---   & ---                 & Ecl \\
 5 & Ecl    & ---   & ---                 & Ecl     \\
 6 & I/II:  & 2.1   & 22.2 (I),  6.9 (II) & Cep(II) \\
 7 & II     & 2.2   & 26.3 (I),  6.7 (II) & Cep(II) \\
 8 & Ecl    & ---   & ---                 & Ecl     \\
 9 & II     & 1.9   & 19.1 (I),  7.7 (II) & Cep(II) \\
10 & II     & 2.6   & 23.2 (I),  6.5 (II) & Cep(II) \\
11 & II     & 2.1   & 21.6 (I),  7.7 (II) & Cep(II) \\
12 & Ecl    & ---   & ---                 & Ecl     \\
13 & Ecl    & ---   & ---                 & Ecl     \\
14 & Ecl    & ---   & ---                 & Ecl     \\
15 & II     & 1.9   & 31.3 (I),  8.0 (II) & Cep(II) \\
16 & Ecl    & ---   & ---                 & Ecl     \\
17 & Ecl    & ---   & ---                 & Ecl     \\
18 & I      & 2.7   &  7.7 (I),  2.3 (II) & Cep(I)  \\
19 & II:    & 0.0   &  9.9 (I),  4.3 (II) & Cep(II) \\
20 & I      & 2.3   &  7.8 (I),  2.4 (II) & Cep(I)  \\
21 & Ecl    & ---   & ---                 & Ecl     \\
22 & Ecl    & ---   & ---                 & Ecl     \\
23 & Ecl    & ---   & ---                 & Ecl     \\
24 & Ecl:   & ---   & ---                 & Ecl     \\
25 & Ecl    & ---   & ---                 & Ecl     \\
26 & I/II:  & 2.4   & 18.7 (I),  5.7 (II) & Cep(II) \\
27 & Ecl    & ---   & ---                 & Ecl     \\
28 & II     & 2.2   & 26.9 (I),  8.3 (II) & Cep(II) \\
29 & I/II:  & 2.3   & 26.6 (I),  9.0 (II) & Cep(II) \\
30 & Ecl    & ---   & ---                 & Ecl     \\
31 & II     & 2.1   & 25.4 (I),  6.4 (II) & Cep(II) \\
32 & I      & 2.4   &  8.3 (I),  2.5 (II) & Cep(I)  \\
33 & II     & 1.1   & 13.0 (I),  4.1 (II) & Cep(II) \\
34 & Cep:   & 1.9  &  3.0 (I),  1.4 (II)  & Cep(?) $\dagger$ \\
35 & Ecl    & ---   & ---                 & Ecl     \\
36 & Ecl    & ---   & ---                 & Ecl     \\
37 & Ecl    & ---   & ---                 & Ecl     \\
38 & Ecl    & ---   & ---                 & Ecl     \\
39 & II     & ---   & ---                 & Cep(II) \\
40 & II     & 2.4   & 28.0 (I),  8.3 (II) & Cep(II) \\
41 & Ecl    & ---   & ---                 & Ecl     \\
42 & Ecl    & ---   & ---                 & Ecl     \\
43 & Ecl    & ---   & ---                 & Ecl     \\
44 & Ecl    & ---   & ---                 & Ecl     \\
45 & II     & 2.1   & 28.6 (I),  8.5 (II) & Cep(II) \\
\hline
\end{tabular}
\end{center}
$\dagger$~The classification of \#3 and \#34 is unclear (see Text).
\end{minipage}
\end{table}

In Fig.~\ref{fig:Fours}
filled circles in black indicate CCEPs and open circles indicate T2Cs.
The two types of Cepheids in the LMC
have reasonably different trends of the Fourier parameters against period.
Thus they are useful for the classification, although there is
a considerable scatter blurring the separation.
The discrimination between the two types can be more robustly done
with estimating their distances than solely based on the light curve shape
(Section~\ref{sec:CepDistance}).

Plus symbols in Fig.~\ref{fig:Fours} indicate eclipsing binaries. Their
$\phi_{21}$ and $\phi_{31}$ values are mostly around $2\pi$
(or equivalently 0) indicating their symmetric variations.
Three objects (\#1, \#4 and \#24) have
$\phi_{31}$ values different from $2\pi$, 
but the amplitudes of the third harmonics are too low.
\#23 ($P=3.26$~d) also has the Fourier parameters unexpected for
an eclipsing binary. This comes from the apparent difference of
the levels outside the eclipsing phases, which however is caused by
the photometric uncertainty due to the crowding effect. An eye inspection of
its light curve suggests that this star is an eclipsing binary.
Light curves of some binaries such as \#1 and \#21 look similar
to those of overtone RR Lyr stars (RRc). However, their amplitudes are
larger than the typical amplitudes of RRc, and furthermore do not show
a decreasing trend with increasing wavelengths which is
a common characteristics of pulsating variables.

Two other objects are indicated by triangles in Fig.~\ref{fig:Fours}.
\#3 ($P=0.265$~d) shows an asymmetric variation typical of pulsating stars.
Also, its amplitude decreases with increasing wavelength, which is
expected for a pulsating star. Its period is at the boundary between
$\delta$~Sct stars and RR~Lyrs in the overtone mode, and
we cannot decide which groups the object belongs to
(also see Section \ref{sec:Classification}).
We consider that \#34 is a Cepheid
but it is unclear to which Cepheid type the object belongs. 
The $\phi_{31}$ of \#34 ($P=2.19$~d) seems to favour
the classification as a CCEP in the overtone mode,
rather than T2Cs, but the $R_{31}$ is much larger than expected.
There is an object classified as an LMC anomalous Cepheid,
OGLE-LMC-ACEP-047, which has the similar Fourier parameters
(see fig.~9 in \citealt{Soszynski-2008b}), although that star itself
shows a slightly different light curve from the majority
of anomalous Cepheids.

\subsection{Reddenings and distances to Cepheids}
\label{sec:CepDistance}

We can also make use of the difference between the absolute magnitudes
of CCEPs and those of T2Cs for the classification.
The estimated distances from the PLRs are very different
depending on the assumed Cepheid population.
Note that the period-colour relations are almost the same for
both types so that a rough estimate of the reddening does not depend
on the classification.

We use the PLRs calibrated with the LMC objects
(Matsunaga, Feast \& Menzies, \citeyear{Matsunaga-2009a})
for T2Cs:
\begin{eqnarray}
J=-2.163 (\pm 0.044) (\log P -1.2) -3.320 (\pm 0.029), \label{eq:PJRII} \\
H=-2.316 (\pm 0.043) (\log P -1.2) -3.720 (\pm 0.028), \label{eq:PHRII} \\
\Ks=-2.278 (\pm 0.047) (\log P -1.2) -3.798 (\pm 0.029). \label{eq:PKRII} 
\end{eqnarray}
Here we assumed the LMC distance modulus 
to be 18.50~mag (\citealt{Benedict-2011}; \citealt{Feast-2012}) 
and the foreground reddening $E_{B-V}$ to be
0.074~mag \citep{Caldwell-1985}.

For the PLRs of CCEPs, we use the calibrating Cepheids
with {\it Hubble Space Telescope} parallaxes \citep{Benedict-2007}.
The $JHK$ magnitudes, on the SAAO system,
listed in \citet{vanLeeuwen-2007} were converted onto
the IRSF/SIRIUS system and further corrected for interstellar extinction
and Lutz-Kelker bias as given in \citet{vanLeeuwen-2007}.
Thus we obtained a linear regression as follows,
\begin{eqnarray}
J=-3.060 (\pm 0.112) (\log P -1.3) -6.219 (\pm 0.057), \label{eq:PJR} \\
H=-3.256 (\pm 0.116) (\log P -1.3) -6.562 (\pm 0.060), \label{eq:PHR} \\
\Ks=-3.295 (\pm 0.121) (\log P -1.3) -6.685 (\pm 0.062), \label{eq:PKR} 
\end{eqnarray}
with scatters of 0.09, 0.10 and 0.10~mag, respectively.
We consider CCEPs only in the fundamental mode
because none of the objects
except a peculiar object (\#34) has a light curve
similar to those of the overtone pulsators.

\begin{figure}
\begin{minipage}{80mm}
\begin{center}
\includegraphics[clip,width=0.9\hsize]{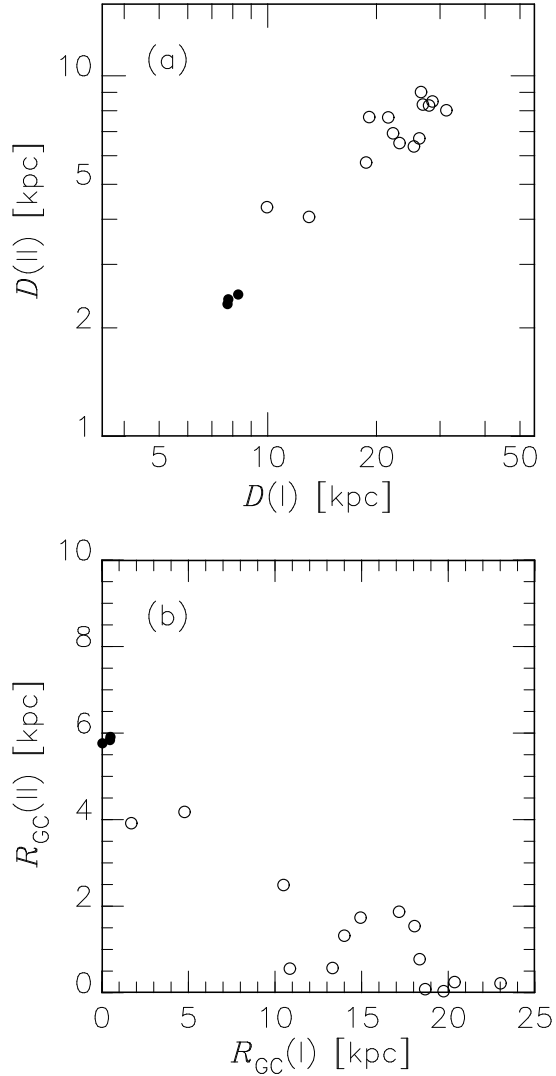}
\end{center}
\caption{
A comparison of predicted distances from the Sun $D$ in panel (a), or
of Galactocentric distances $R_{\rm GC}$ in panel (b),
assuming two types of Cepheids is done for the variables which are
considered to be CCEPs or T2Cs. $D$(I) and $R_{\rm GC}$(I) indicate
the values assuming the type of CCEP, and $D$(II) and $R_{\rm GC}$(II)
assuming the type of T2C. The filled and open circles indicate CCEPs
and T2Cs respectively.
\label{fig:compDMs}}
\end{minipage}
\end{figure}

For each Cepheid candidate, the distance and extinction are tentatively
derived using the PLRs of both types of
Cepheid (Table~\ref{tab:classification}).
As we discussed in Paper~I, an estimate of $(\mu_0, \AK)$ is possible
with a pair of two-band photometry, and three estimates can be obtained
with $\JHK$ magnitudes.
The reddening law in $\JHK$ is taken from \citet{Nishiyama-2006a}.
The panel (a) of Fig.~\ref{fig:compDMs} compares the distances from the Sun
assuming that the variables are CCEPs, $D$(I),
with those assuming that they are T2Cs, $D$(II).
For example, the objects with $D{\rm (II)}\sim 8$~kpc would be
further than 20~kpc if assumed to be CCEPs.
It is almost certain that such stars are T2Cs in the Galactic bulge
rather than CCEPs far behind the GC, especially when their extinctions are
not larger than the values expected at the distance of the GC.

In addition, there is a constraint on the distribution of T2Cs;
they are concentrated to the Galactic bulge. Paper~I showed
that short-period Miras ($P\leq 350$~d) found in the same survey
are strongly concentrated to 
the distance of the GC ($\sim 8.24$~kpc) and also that
they suffer from interstellar extinctions larger than $\sim 2$~mag in $\Ks$.
One can assume as a first approximation  
that T2Cs are distributed in the same manner because
such short-period Miras are considered to be as old as T2Cs.
Panel (b) of Fig.~\ref{fig:compDMs} compares the Galactocentric distances
under the two assumptions, $R_{\rm GC}$(I) and $R_{\rm GC}$(II)
(here we assumed the GC distance is 8.24~kpc, Paper~I).
The bulk of the open circles are concentrated towards small $R_{\rm GC}$(II),
but they would be significantly further than the GC
if assumed to be CCEPs. The extinction $\AK$ are estimated to be 2--2.5~mag
for these objects, regardless of the Cepheid type. This is the approximate
range of values expected for objects at the GC distance. 
Thus they are considered to be T2Cs in the Galactic bulge.
Two objects with $R_{\rm GC}{\rm (II)}\sim 4$~kpc fall
at the intermediate range in Fig.~\ref{fig:compDMs} (\#19 and \#33).
However, their small extinctions 
strongly suggest that they are relatively close T2Cs
rather than CCEPs further than the GC.
In contrast, three objects are found to be CCEPs as we reported in Paper~II.

The periods of six T2Cs are longer than 20~d. From the work on the T2Cs
in the Magellanic Clouds, it is known that such long-period T2Cs show
a large scatter in the period-magnitude diagrams and may be systematically
brighter than the PLR obtained for the shorter-period T2Cs, BL~Her
and W~Vir types \citep{Matsunaga-2009a}.
The scatter, however, is not so large as to change the classification.

According to the variable type determined here, the distance moduli
and extinctions are derived and listed in Table~\ref{tab:muAK}.
Mean estimates of the ($\mu_0, \AK$),
whenever available, are also listed in Table~\ref{tab:muAK} and they are
used in the following discussions. The estimates from the different pairs
of filters agree reasonably well with each other,
except the case of \#2 ($P=7.46$~d) whose photometry is uncertain
due to the effect of crowding. Inconsistent $\AK$ estimates from the $JH$
and $H\Ks$ pairs occur if the measured colours are not in accordance with
the sum of the intrinsic colours and the reddening vector.
Such inconsistency can happen when blue and red stars are merged
in the line of sight (see the discussion in section~4.2 of Paper~I).

\begin{table*}
\begin{minipage}{170mm}
\caption{
Estimated distance moduli and extinctions for Cepheids.
The pairs of two filters used to estimate the values are given as superscripts, while the mean values of available estimates for each star are given in the last columns.
Types of Cepheids (I=classical Cepheids, II=type II Cepheids) and periods are also indicated.
\label{tab:muAK}}
\begin{center}
\begin{tabular}{rcrrrrrrrrr}
\hline
\multicolumn{1}{c}{No.} & Type & $\log P$ & $\mu_0^{HK}$ & $A_{\Ks}^{HK}$ & $\mu_0^{JK}$ & $A_{\Ks}^{JK}$ & $\mu_0^{JH}$ & $A_{\Ks}^{JH}$ & $\mu_0^{\rm Mean}$ & $A_{\Ks}^{\rm Mean}$ \\
\hline
 2 & II & 0.873 & 14.17: & 1.72: & 14.66: & 1.23: & 15.49: & 0.96: & 14.77: & 1.30: \\
 6 & II & 1.179 & 14.23 & 2.09 &  \multicolumn{1}{c}{---}  &  \multicolumn{1}{c}{---}  &  \multicolumn{1}{c}{---}  &  \multicolumn{1}{c}{---}  & 14.23 & 2.09 \\
 7 & II & 1.718 & 14.12 & 2.24 & 14.12 & 2.24 & 14.12 & 2.24 & 14.12 & 2.24 \\
 9 & II & 0.676 & 14.47 & 1.87 &  \multicolumn{1}{c}{---}  &  \multicolumn{1}{c}{---}  &  \multicolumn{1}{c}{---}  &  \multicolumn{1}{c}{---}  & 14.47 & 1.87 \\
10 & II & 1.382 & 14.11 & 2.62 &  \multicolumn{1}{c}{---}  &  \multicolumn{1}{c}{---}  &  \multicolumn{1}{c}{---}  &  \multicolumn{1}{c}{---}  & 14.11 & 2.62 \\
11 & II & 0.918 & 14.41 & 2.09 &  \multicolumn{1}{c}{---}  &  \multicolumn{1}{c}{---}  &  \multicolumn{1}{c}{---}  &  \multicolumn{1}{c}{---}  & 14.41 & 2.09 \\
15 & II & 1.703 & 14.18 & 2.06 & 14.32 & 1.92 & 14.56 & 1.84 & 14.35 & 1.94 \\
18 & I  & 1.357 & 14.50 & 2.68 & 14.43 & 2.75 & 14.32 & 2.79 & 14.42 & 2.74 \\
19 & II & 0.560 & 13.07 & 0.04 & 13.18 & -0.07 & 13.36 & -0.13 & 13.20 & -0.05 \\
20 & I  & 1.300 & 14.53 & 2.32 & 14.50 & 2.35 & 14.45 & 2.37 & 14.49 & 2.35 \\
26 & II & 1.207 & 13.93 & 2.38 &  \multicolumn{1}{c}{---}  &  \multicolumn{1}{c}{---}  &  \multicolumn{1}{c}{---}  &  \multicolumn{1}{c}{---}  & 13.93 & 2.38 \\
28 & II & 1.192 & 14.60 & 2.18 &  \multicolumn{1}{c}{---}  &  \multicolumn{1}{c}{---}  &  \multicolumn{1}{c}{---}  &  \multicolumn{1}{c}{---}  & 14.60 & 2.18 \\
29 & II & 1.011 & 14.60 & 2.31 &  \multicolumn{1}{c}{---}  &  \multicolumn{1}{c}{---}  &  \multicolumn{1}{c}{---}  &  \multicolumn{1}{c}{---}  & 14.60 & 2.31 \\
31 & II & 1.603 & 13.97 & 2.13 &  \multicolumn{1}{c}{---}  &  \multicolumn{1}{c}{---}  &  \multicolumn{1}{c}{---}  &  \multicolumn{1}{c}{---}  & 13.97 & 2.13 \\
32 & I  & 1.372 & 14.69 & 2.36 & 14.57 & 2.48 & 14.38 & 2.54 & 14.55 & 2.46 \\
33 & II & 1.278 & 13.05 & 1.04 & 13.04 & 1.05 & 13.02 & 1.06 & 13.04 & 1.05 \\
40 & II & 1.279 & 14.63 & 2.44 &  \multicolumn{1}{c}{---}  &  \multicolumn{1}{c}{---}  &  \multicolumn{1}{c}{---}  &  \multicolumn{1}{c}{---}  & 14.63 & 2.44 \\
45 & II & 1.387 & 14.51 & 2.10 & 14.55 & 2.06 & 14.62 & 2.04 & 14.56 & 2.07 \\
\hline
\end{tabular}
\end{center}
\end{minipage}
\end{table*}

The object \#39 ($P=30.9$~d) was detected only in the $\Ks$-band, so that 
the distance and extinction cannot be obtained. This star is much fainter
than the three CCEPs in spite of the fact that the period is longer than theirs.
If this star is a CCEP at the distance of the GC, the extinction $\AK$
should be as large as 5.5~mag. In contrast, a T2C with $\AK \sim 2.8$ would 
be consistent with the observed $\Ks$ magnitude and the faintness in 
$J$ and $H$. We conclude that this star is a T2C in the Galactic bulge.

\subsection{Summary of the classification}
\label{sec:Classification}
The previous subsections show that most of the variables can be reasonably 
classified. The adopted types are indicated in the last columns of
Table~\ref{tab:catalogue} and \ref{tab:classification}.
About half, 24, of the objects are classified as eclipsing binaries.
Three are CCEPs and 16 are T2Cs. \#3 is a pulsating star
with a short period, 0.265~d, and falls in the period range between
RR~Lyr and $\delta$~Sct stars. The classification of \#34 is uncertain.

\begin{figure*}
\begin{tabular}{cc}
\begin{minipage}{80mm}
\begin{center}
\includegraphics[clip,width=0.98\hsize]{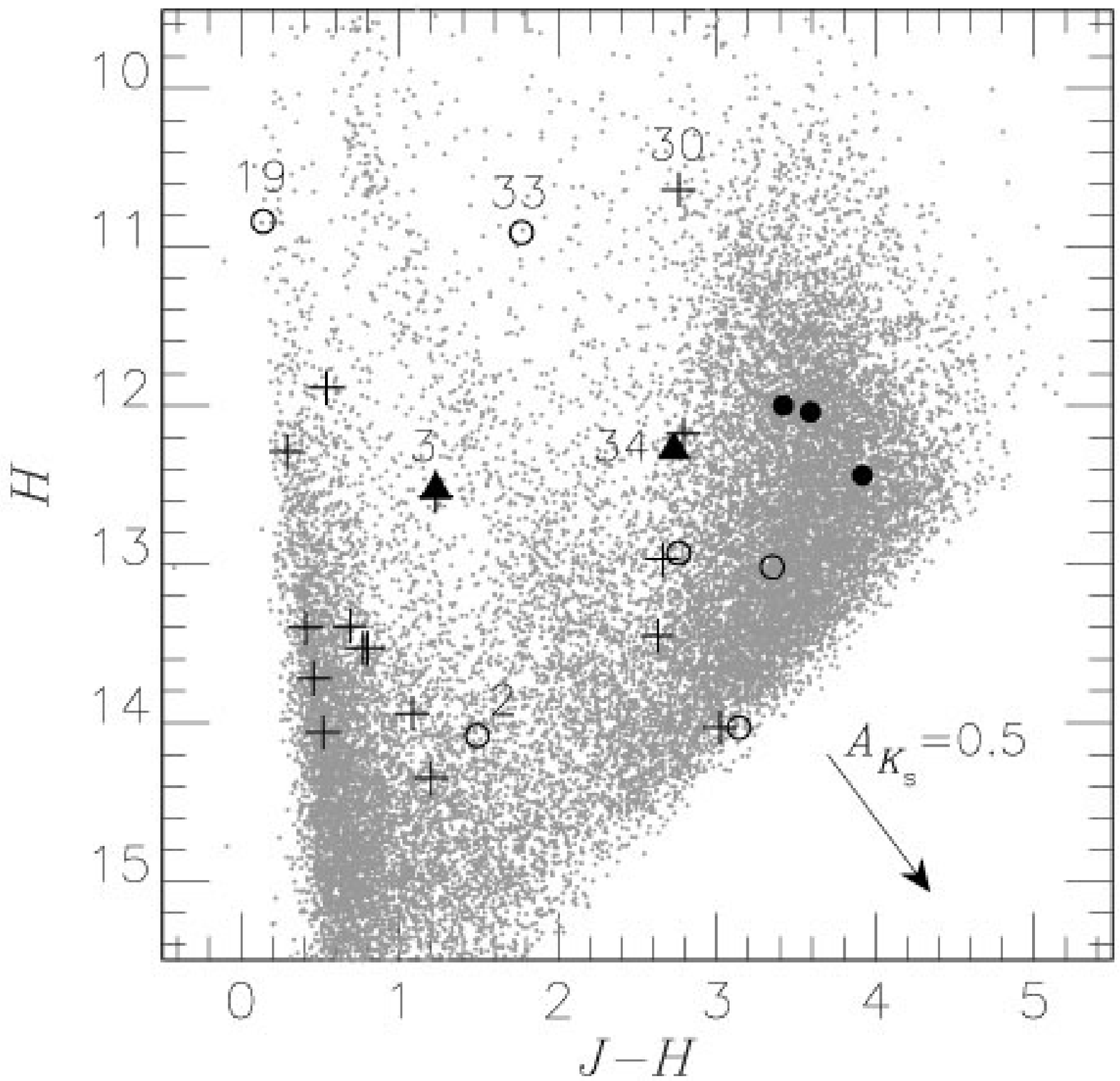}
\end{center}
\end{minipage}
\begin{minipage}{80mm}
\begin{center}
\includegraphics[clip,width=0.98\hsize]{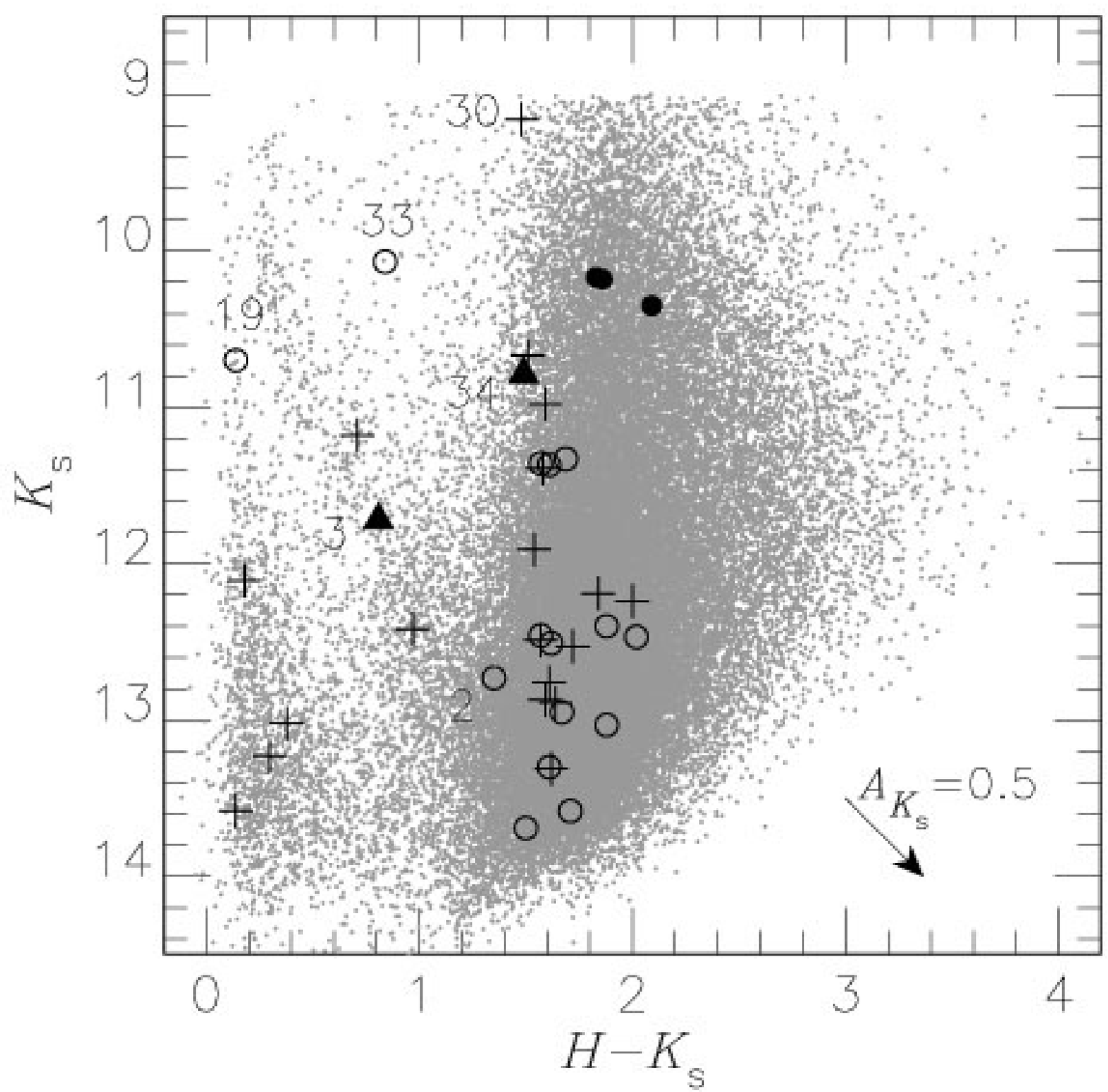}
\end{center}
\end{minipage}
\end{tabular}
\caption{
Colour-magnitude diagrams: $J-H$ vs $H$ (left) and $H-\Ks$ vs $\Ks$ (right).
The short-period variables
are indicated by symbols depending on the classified variable types:
eclipsing variables by plus symbols, CCEPs by filled circles,
T2Cs by open circles and the other variables by triangles,
while grey dots plot other objects.
The numbered objects are discussed in the text.
The arrow indicates the reddening vector corresponding to $\AK=0.5$~mag.
\label{fig:CMDs}}
\end{figure*}

Fig.~\ref{fig:CMDs} shows colour-magnitude diagrams for our catalogued
variables and the other sources we detected in the survey.
Open circles indicate T2Cs.
The foreground T2Cs, \#19 and \#33, are relatively blue and bright.
The $(J-H)$ colour of a faint T2C, \#2, is blue
but its photometry was affected by the crowding.
The other T2Cs are reddened and lie on the broadened giant branch
of the Galactic bulge. Three CCEPs indicated by filled circles are located
close to each other on the colour-magnitude diagrams;
they are significantly reddened but relatively bright.

Eclipsing binaries, plus symbols in Fig.~\ref{fig:CMDs}, are 
separated roughly into two groups, around the foreground main sequence or
on the giant branch of the bulge. A few points exist in the intermediate 
colour range, \#8, \#36 and \#41, but their colours are affected
by the crowding (see their $M$flag in Table~\ref{tab:catalogue}). 
The colours of the redder group suggest that they have large interstellar
extinction and are thus distant and likely in the GC region.
Excluding those affected by the crowding, this group
includes \#17, \#21, \#25, \#27, \#30, \#37,
\#38, \#42 and \#43. They tend to have longer orbital periods than 
the bluer binaries.
The brightest of the reddened binaries, \#30 ($P=26.8$~d), is
of particular interest. It was reported as an O-type supergiant
located near the GC \citep{Mauerhan-2010},
but we find that it is a binary system.
Furthermore, its asymmetric light curve suggests that the system
has an eccentric orbit.
Since the other reddened binaries may well be at the distance of the GC,
they are also interesting objects for further study.

The triangle for \#3 falls near 
the diagonal sequence of the red clump giants
in the disk \citep{Lucas-2008},
where the RR~Lyr and $\delta$~Sct stars in the foreground
are roughly expected. On the other hand, \#34 is highly
reddened and relatively bright,
although the images in the $H$ and $\Ks$ bands indicate that
the photometry may be affected by crowding.
Its distance would be 3.6~kpc if it were an overtone CCEP,
and the distance would be smaller otherwise.
Therefore it is much closer than the GC, and yet the extinction is
quite high, $\AK\sim 2$~mag. These values may be subject to the uncertainty
due to the crowding, but 
it would not change the conclusion that
this object is in the foreground of the GC. The nature of the star remains
to be investigated.

\section{Discussion}

\subsection{Samples of Type II Cepheids}

\begin{figure}
\begin{minipage}{80mm}
\begin{center}
\includegraphics[clip,width=0.9\hsize]{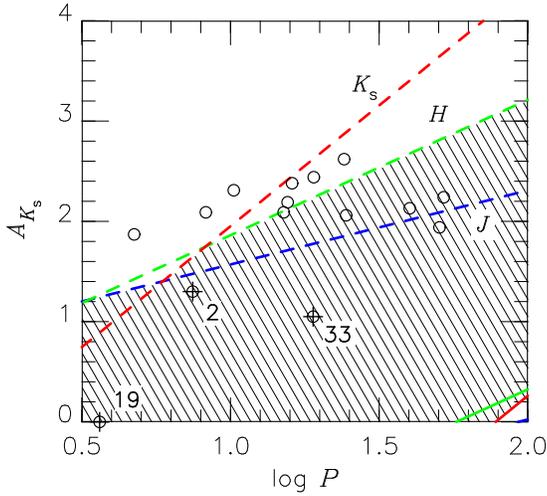}
\end{center}
\caption{
A schematic diagram to show the range,
in the parameter space of $(\log P, \AK)$,
where we can detect T2Cs located at the GC distance. Dashed lines
show the detection limits for $\JHK$, while solid lines indicate the saturation
limits though scarcely relevant.
Blue green and red lines correspond to the limits
in $J$, $H$ and $\Ks$, respectively.
The shaded region illustrates
the range for which the T2Cs are detected in two or more filters.
The open circles are plotted for the $(\log P, \AK)$ values
of the T2Cs we discovered in this work.
Star \#2 (uncertain photometry) and
the stars \#19 and \#33 (foreground) are indicated.
\label{fig:GCCepRange2}}
\end{minipage}
\end{figure}

Our catalogue includes 16 T2Cs; 
14 are located in the bulge and two in the foreground (\#19 and \#33).
In the following discussion, we consider 
14 objects as our sample of T2Cs
in the bulge unless otherwise mentioned. 
We found few T2Cs with short period ($P<5$~d).
In fact our survey was not deep enough to detect such objects.
The PLR enables us to tell if the detection
limit is deep enough to detect a Cepheid
with a given period, foreground extinction and distance.
Fig.~\ref{fig:GCCepRange2} illustrates the range
in the parameter plane of $(\log P, \AK)$ where we should be able to
detect T2Cs at the distance of the GC.
For example, a T2C with $\log P=1$ could not be detected 
if the foreground extinction is larger than $\AK=2$~mag.
Considering that the majority of the objects near the GC
are reddened more strongly, our survey is far from complete.
In Fig.~\ref{fig:GCCepRange2},
several of the detected T2Cs seem to be beyond the detection limit in
the $H$ band or even in the $\Ks$ band.
This happens because the limiting magnitude 
depends on the crowdedness which varies within our survey region.
Our survey region includes
the extremely crowded region near Sgr~A$^*$
and also sparse regions towards dark cloud lanes.

Recently, \citet{Soszynski-2011} found a rich population of T2Cs
in the outer bulge towards low-extinction regions
away from the Galactic plane.
We combined their catalogue with the 2MASS near-infrared catalogue
\citep{Skrutskie-2006} with the tolerance radius of 1~arcsec.
There are 156 BL~Her, 128 W~Vir and 51 RV~Tau objects
in the OGLE-III catalogue, and we found 97, 117 and 49 counterparts
for the three types of T2Cs respectively
(263 in total, Table~\ref{tab:OGLE2MASS}).
Because of their faintness, a significant fraction of the BL~Her stars
were not detected in 2MASS. In addition, the 2MASS catalogue indicates 
that photometric accuracies for quite a few objects are limited
because of confusion or other reasons.
Considering the quality flag, the blend flag
and the confusion flag (Qflag, Bflag and Cflag in Table~\ref{tab:OGLE2MASS}), 
there remain 166, 138 and 138 measurements in the $\JHK$ bands
with good photometric quality. We discriminate these "good" magnitudes
from the others below.
In addition, we use the $I$-band light curves obtained by the OGLE-III survey
to make phase corrections to convert the single-epoch 2MASS magnitudes
into mean magnitudes. The same method was described and used in
Matsunaga et~al.~(\citeyear{Matsunaga-2009a}, \citeyear{Matsunaga-2011a}).
Table~\ref{tab:OGLE2MASS} lists 
the size of the correction, $\Delta_{\phi}$, which is to be added
to the 2MASS magnitudes for each object.
Some $I$-band light curves show a large scatter, and we do not apply
the phase correction for those T2Cs (mainly RV~Tau stars,
with $\Delta_{\phi}=99.999$ in Table~\ref{tab:OGLE2MASS}).

\begin{table*}
\begin{minipage}{170mm}
\caption{
2MASS counterparts for the OGLE-III type II Cepheids in the bulge.
The table lists the period $P$ and the 2MASS $\JHK$ magnitudes as well as
the quality flags: Qflag=the photometric quality flag, Bflag=the blend flag,
and Cflag=the confusion flag \citep{Skrutskie-2006}.
In the last column listed are the phase correction $\Delta_{\phi}$
we obtained with the OGLE $I$-band light curves (see the text).
This is the first 10 lines of the full catalogue which will be available
in the online version of the article (see Supporting Information).
\label{tab:OGLE2MASS}}
\begin{center}
\begin{tabular}{crcrrrcccr}
\hline
OGLE~ID & \multicolumn{1}{c}{$P$} & 2MASS~ID & \multicolumn{1}{c}{$J$} & \multicolumn{1}{c}{$H$} & \multicolumn{1}{c}{$\Ks$} & Qflag & Bflag & Cflag & \multicolumn{1}{c}{$\Delta_{\phi}$} \\
 & \multicolumn{1}{c}{(days)} & & \multicolumn{1}{c}{(mag)} & \multicolumn{1}{c}{(mag)} & \multicolumn{1}{c}{(mag)} & & & & \multicolumn{1}{c}{(mag)} \\
\hline
OGLE-BLG-T2CEP-001 & 3.9983508 &17052035-3228176 &12.904 &12.412 &12.273 &AAA &111 &000 &$ 0.160$ \\
OGLE-BLG-T2CEP-002 & 2.2684194 &17061499-3301275 &13.130 &12.713 &12.587 &AAA &111 &000 &$-0.050$ \\
OGLE-BLG-T2CEP-003 & 1.4844493 &17084014-3254104 &13.826 &13.457 &13.276 &AAA &111 &00c &$ 0.295$ \\
OGLE-BLG-T2CEP-004 & 1.2118999 &17131083-2905453 &13.817 &13.355 &13.248 &AAA &111 &000 &$ 0.052$ \\
OGLE-BLG-T2CEP-006 & 7.6379292 &17142541-2846465 &12.213 &11.863 &11.682 &AAA &111 &000 &$ 0.266$ \\
OGLE-BLG-T2CEP-007 & 1.8173297 &17235478-2902378 &14.293 &13.607 &13.426 &AAA &111 &000 &$-0.200$ \\
OGLE-BLG-T2CEP-008 & 1.1829551 &17242093-2755493 &14.573 &99.999 &99.999 &AUU &200 &c00 &$ 0.091$ \\
OGLE-BLG-T2CEP-009 & 1.8960106 &17242227-2927352 &14.218 &13.539 &13.245 &AAA &112 &0dc &$-0.159$ \\
OGLE-BLG-T2CEP-010 & 1.9146495 &17270554-2536015 &13.671 &13.143 &12.989 &AAA &111 &000 &$ 0.138$ \\
OGLE-BLG-T2CEP-011 &15.3886022 &17271765-2538234 &12.006 &11.208 &10.995 &AAA &111 &000 &$-0.300$ \\
\hline
\end{tabular}
\end{center}
\end{minipage}
\end{table*}

\subsection{Period distribution and surface density}

\begin{figure}
\begin{minipage}{80mm}
\begin{center}
\includegraphics[clip,width=0.98\hsize]{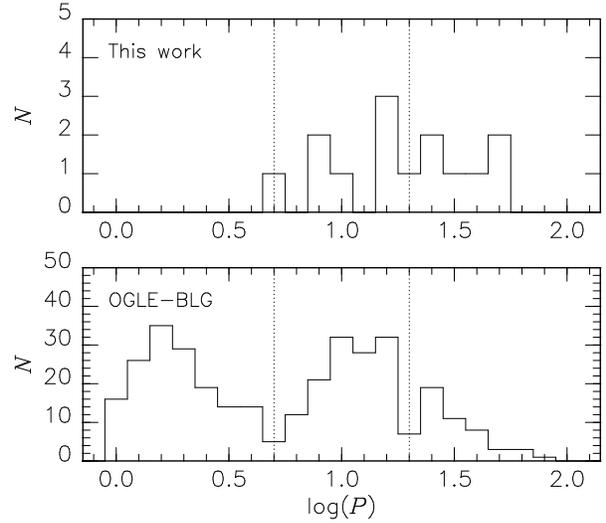}
\end{center}
\caption{
Histograms of periods for the T2Cs in the bulge:
the top panel for our sample and the bottom for the OGLE-III sample
\citep{Soszynski-2011}. Vertical lines show the thresholds,
5 and 20~d, used to divide BL~Her, W~Vir and RV~Tau variables.
\label{fig:PhistT2C}}
\end{minipage}
\end{figure}

The period distributions of our T2Cs and the OGLE-III sample
are shown in Fig.~\ref{fig:PhistT2C}.
Most of our sample have long periods ($P>10$~d). This bias is
caused by the detection limit of our survey as mentioned above.
In contrast, with more than 300 T2Cs,
the OGLE-III sample clearly shows the distinct groups
of BL~Her, W~Vir, RV~Tau stars. Such a feature is well seen in
the T2C samples of the Magellanic Clouds but not in that of globular clusters
(see fig.~5 in \citealt{Matsunaga-2011a}).

In addition, there are significantly more W~Vir stars than RV~Tau stars
and the periods of W~Vir stars show a broad, or even two distinct,
peak(s), both of which are similar to the case of the LMC T2Cs.
The number of BL~Her stars is even larger than W~Vir stars,
i.e.~$N_{\rm WV}/N_{\rm BL}=0.82\pm 0.14$ (error from Poisson noise).
This ratio falls between the case of the LMC (1.25) and the SMC (0.6)
given in \citet{Matsunaga-2011a}.
The reason for these variations in the T2C populations is presumably
related to age and/or metallicity, but the lack of theoretical models
for T2Cs prevents us from further discussion.

\begin{figure}
\begin{minipage}{80mm}
\begin{center}
\includegraphics[clip,width=0.98\hsize]{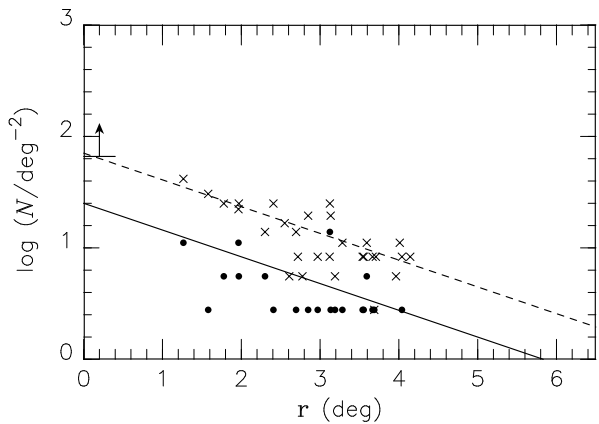}
\end{center}
\caption{
Density profile of T2Cs in the bulge.
Crosses and filled circles indicate the surface densities [deg$^{-2}$]
of all T2Cs and those with $P>15$~d \citep{Soszynski-2011}
in the OGLE-III fields with $|l|<2^\circ$ and $|b|<4^\circ$. 
The lower limit of the density of T2Cs with $P>15$~d towards the GC
is indicated by the arrow.
The exponential law indicated by the straight lines
gives reasonable fits to the OGLE-III points.
\label{fig:density_T2C}}
\end{minipage}
\end{figure}

\begin{figure}
\begin{minipage}{80mm}
\begin{center}
\includegraphics[clip,width=0.98\hsize]{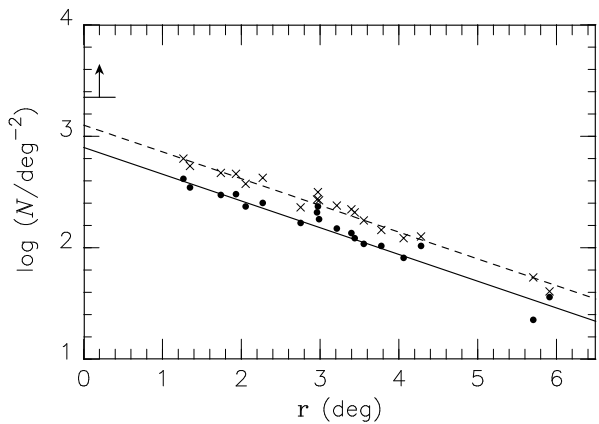}
\end{center}
\caption{
Same as Fig.~\ref{fig:density_T2C}, but for Miras.
Crosses and filled circles indicate the surface densities [deg$^{-2}$]
of all Miras with known periods and those with $P<350$~d \citep{Matsunaga-2009b}
in the OGLE-II fields with $|l|<2^\circ$ and $|b|<4^\circ$. 
The lower limit of the density of Miras with $P<350$~d towards the GC
is indicated by the arrow. 
The exponential law indicated by the straight lines
gives reasonable fits to the OGLE-II points.
\label{fig:density_Mira}}
\end{minipage}
\end{figure}

It is of interest to examine the surface density of T2Cs in the bulge.
We detected 11 T2Cs with $P>15$~d,
which leads to the density of 66~deg$^{-2}$
considering the area of our survey towards the GC (1/6~deg$^{-2}$).
However, our survey was not complete even for the relatively long-period T2Cs
because of thick dark nebulae
(Fig.~\ref{fig:GCCepRange2}), and the above density 
is an underestimate thus indicated by 
the arrow in Fig.~\ref{fig:density_T2C}.
For the outer bulge region, we obtained the surface density of T2Cs for
each OGLE-III region.
Fig.~\ref{fig:density_T2C} plots the surface densities of
the OGLE-III T2Cs with $P>15$~d (filled circles) and all T2Cs (crosses)
in each field against the angular distance from the GC.
We consider only the fields in the range of $|l|<2^\circ$ and $|b|<4^\circ$ 
where the density is high enough.
The profile in Fig.~\ref{fig:density_T2C} shows, in effect,
the variation along the minor axis.
In addition, Fig.~\ref{fig:density_Mira} shows a similar plot
of the density profile
for Miras. \citet{Matsunaga-2011b} found 547 Miras with period determined
in the same IRSF survey field, among which 251 objects have periods less than
350~d. Whilst the Miras have a broad range of age
(from $\sim 10$~Gyr to 1~Gyr or even younger),
such short-period Miras are found in globular clusters 
\citep{Frogel-1998} 
and considered to belong to the old stellar population.
The number of the short-period Miras towards the GC field corresponds to
a surface density of 2200~deg$^{-2}$. 
The surface densities for the outer region were obtained
using the catalogue of the OGLE-II Miras compiled by
Matsunaga, Fukushi \& Nakada~(2005). 
In Fig.~\ref{fig:density_Mira},
the density profile for the OGLE-II Miras is well represented
by the exponential law, $N\sim \exp (-0.24 r)$.
This exponential fits the OGLE-II points better than
a de Voucouleurs law or a Sersic law, $\log N \sim R^{1/n}$, with $n=2$. 
Note that the exponential and the de Voucouleurs law correspond
to the Sersic law with $n=1$ and $n=4$ respectively.
In contrast, the lower limit inferred by our sample is higher than
the exponential law predicted by the OGLE-II Miras. This excess agrees
with the idea that an additional population of Miras exist in
the nuclear bulge, the disk-like system within $\sim 200$~pc.
Although the density profile for T2Cs is uncertain due to the small number,
our result on the T2C distribution also suggests that
the nuclear bulge holds an additional group of T2Cs in the central region.

\subsection{The distance to the GC}

There have been a considerable number of estimates of the distance
to the Galactic Centre based on stellar distance indicators.
Many of these rely on data from the general region of the Galactic bulge
and may, to a greater or lesser degree, be affected by the bar-like and
other structure of the bulge. In this section, we concentrate on data
obtained in the areas close to the Centre which should be
free of any such effects.

\begin{figure}
\begin{minipage}{80mm}
\begin{center}
\includegraphics[clip,width=0.98\hsize]{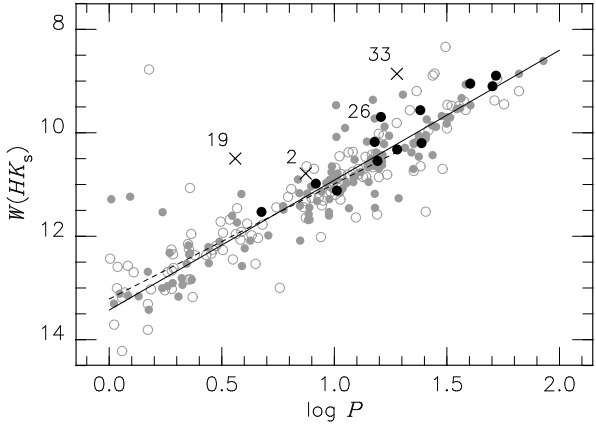}
\end{center}
\caption{
The $\log P$-$W(H\Ks)$ relation for T2Cs in the bulge.
Black filled circles indicate our T2Cs, but
crosses indicate two foreground stars (\#19 and \#33) and
\#2 with uncertain photometric result.
The stars in the OGLE III catalogue \citep{Soszynski-2011}
are also plotted as the grey symbols, of which the open circles 
are used for the objects with uncertain 2MASS photometry or
no phase correction applied (see text)
and filled circles for the others.
Dashed and filled relations plot the $\log P$-$W(H\Ks)$ relations for
the LMC T2Cs \citep{Matsunaga-2009a} and those in globular clusters
\citep{Matsunaga-2006}, respectively, but after the distance effect
corrected assuming $\mu _0$(LMC)=18.50~mag and $\mu _0$(bulge)=14.50~mag. 
\label{fig:PW}}
\end{minipage}
\end{figure}

In view of the importance of reddening in this region
we use a reddening free PLR in $H$ and $\Ks$,
\begin{equation}
W(H\Ks) = \Ks - 1.44 (H-\Ks),
\end{equation}
where the coefficient is taken from the extinction law
found by \citet{Nishiyama-2006a}. 
As a preliminary we compare our T2C results with those from the OGLE-III
survey in the general bulge. Fig.~\ref{fig:PW} plots $W(H\Ks)$ 
against the period for both our sample and the OGLE-III sample
with $H$ and $\Ks$ magnitudes.
Of our sample, 13 are indicated by the filled circles, whereas 
the crosses indicate two foreground stars (\#19 and \#33) and
\#2 with uncertain photometry. 
The grey symbols show the OGLE-III objects (filled circles
for those with good $H/\Ks$ magnitudes and open circles for others).
The linear relations drawn in the Fig.~\ref{fig:PW} are obtained with
the T2Cs in globular clusters (filled line; \citealt{Matsunaga-2006})
and those in the LMC (dashed line; \citealt{Matsunaga-2009a})
but with a shift considering the approximate distance moduli
of the LMC (18.50~mag) and the bulge (14.50~mag).
Most of our T2Cs except the crosses (\#2, \#19 and \#33) lie close together
with the samples of the OGLE-III catalogue
in the vicinity of the GC, $\sim 8$~kpc (Fig.~\ref{fig:compDMs}).
In contrast, \#26 ($\log P = 1.21$) seems brighter
than the relation for other stars.
If it is a normal T2C it lies in the foreground of the bulge
while it may be a peculiar W~Vir stars brighter than regular W~Vir stars.
Several objects in the OGLE-III are also brighter than the others,
and their nature needs to be investigated.

We now concentrate on our T2Cs, which are in the vicinity of the GC,
and whose periods fall within the range of W~Vir stars.
Previous work found that the PLRs of BL~Her/RV~Tau stars
may be different between different galaxies
(Matsunaga et~al. \citeyear{Matsunaga-2009a}, \citeyear{Matsunaga-2011a};
\citealt{Soszynski-2011}), although we did not confirm significant deviation
of our T2C samples from the PLR of those in globular clusters
(Fig.~\ref{fig:PW}).
We obtained the average modulus of 
$\mu_0 = 14.38\pm 0.13$~mag, based on 5 W~Vir stars
($5\geq P\geq 20$~d) excluding \#26,
under the assumption of $\muLMC = 18.50$~mag.
The errors in the above estimates account just for statistical errors,
and we need to consider systematic uncertainties.
Our estimates are affected by errors in the extinction law and
the LMC distance as well as the possible 
population effect on the PLR.
We adopt an uncertainty of 0.05~mag for the LMC modulus and
0.07~mag for the adopted reddening law as we did in Paper~I
for the GC Miras. The results of \citet{Matsunaga-2006}
and \citet{Matsunaga-2011a} suggest that any population effect
on the PLR of T2Cs (W~Vir stars) is small. Nevertheless, to be conservative
we adopt an uncertainty of 0.07~mag for this.
Considering these errors and the above estimates,
the current sample of T2Cs results in an estimate of
the GC distance modulus to be $14.38 \pm 0.17$~mag. 
There is a further uncertainty due, as discussed above, to our detection limit.
This might result in the modulus being slightly
underestimated\footnote{47 OGLE-III T2Cs in the same period range give
a modulus of $14.40\pm 0.05$ (internal error).}.
With the same survey data, we obtained the distances to
Miras (Paper~I) and CCEPs (Paper~II).
Adopting $\muLMC = 18.50$~mag, the average of the distances to Miras
gives $\muGC = 14.63 \pm 0.11$~mag.
On the other hand, the calibration of CCEPs are based on the nearby calibrators
and the average of the three distances leads to $\muGC = 14.49 \pm 0.12$~mag.
The error budgets for these estimates are discussed in Papers~I and II.

$\JHK$ observations of red clump stars in the region around the Galactic Centre
were obtained by \citet{Nishiyama-2006b} using the IRSF/SIRIUS. 
Recently, Laney, Joner \& Pietrzy\'{n}ski (2012) 
obtained new high-precision $\JHK$ magnitudes of red clump giants
with the Hipparcos parallaxes, which gives a new calibration of the red clump.
\citet{Nishiyama-2006b} adopted $(H-\Ks)_0=0.07$ and $\Ks=-1.59$
from theoretical isochrones by
(Bonatto, Bica \& Girardi, 2004), 
whereas \citet{Laney-2012} obtained $(H-\Ks)_0=0.123$ and $\Ks=-1.613$.
Using this new calibration leads to $\muGC = 14.53 \pm 0.10$~mag,
that is $8.05 \pm 0.37$~kpc, without any population effect taken into account.
There is a large scatter in metallicities of red clump giants
in the bulge and the median metallicity seems slightly higher than
the solar abundance \citep{Hill-2011}. 
The error, adopted from \citet{Nishiyama-2006b}, includes
and is affected by the uncertainty of a possible population effect.

\begin{figure}
\begin{minipage}{80mm}
\begin{center}
\includegraphics[clip,width=0.98\hsize]{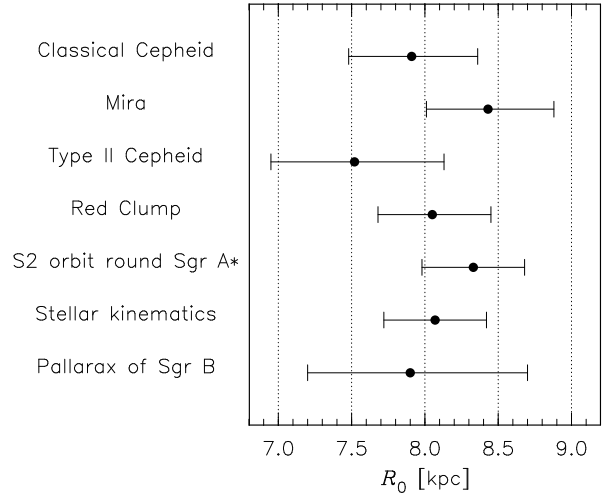}
\end{center}
\caption{
Estimates of the distance $R_0$ to the GC based on various methods
(see the text for references). Error bars include the statistical
and systematic uncertainties.
\label{fig:GCDMs}}
\end{minipage}
\end{figure}

These estimates based on near-IR data of stellar distance indicators in areas
close to the Centre are compared with the results from kinematic methods
in Fig.~\ref{fig:GCDMs}. These latter methods are:
the Kepler rotation of the star S2 around Sgr~A$^*$ \citep{Gillessen-2009},
the statistical parallax method applied to the central stellar cluster
\citep{Trippe-2008} and the parallax of Sgr~B \citep{Reid-2009}.
The photometric and kinematic determinations are in satisfactory
agreement and indicate a value of $R_0$, close to 8.0~kpc. 
The uncertainty in the reddening law is the dominant remaining error
for the photometric distances discussed here. Thus the agreement of
the photometric and kinematic results lends support to
the reddening law of \citet{Nishiyama-2006a}.
For a typical value of $E_{H-\Ks}=1.8$, for instance,
the Nishiyama value of $\AK$ is
2.5~mag whereas the \citet{Rieke-1985} law gives 3.2~mag and
would lead to an unacceptably small value of $R_0$(GC).

\section{Summary\label{Summary}}

Through our near-IR survey of stellar variability towards the GC,
45 short-period variables have been discovered.
Their light curves are investigated to determine the variable types,
and for the Cepheid candidates their distances and foreground extinctions
are also considered based on the PLRs. Most of the objects
are reasonably classified: three CCEPs, 16 T2Cs, 24 eclipsing binaries,
and two others. The numbers of T2Cs and short-period
Miras in our survey region
are higher than the surface density following the exponential law 
which fits the distribution of T2Cs and Miras in the outer bulge.
This strongly suggests that the nuclear bulge hosts a significant
population of old stars ($\geq 10$~Gyr).
We also discuss the distance to the Galactic Centre based on
stellar distance indicators in the central region. These are
insensitive to problems associated with the three dimensional structure
of the bulge which may affect other determinations. Our main result is
close to 8~kpc and agrees well with kinematic estimates.
Since the photometric results are rather sensitive to the infrared
reddening law, the result give support to the reddening law
of \citet{Nishiyama-2006a} which we adopted.

\section*{Acknowledgments}
We thank the IRSF/SIRIUS team and the staff of South African Astronomical
Observatory (SAAO) for their support during our near-IR observations.
The IRSF/SIRIUS project was initiated and supported by Nagoya University,
National Astronomical Observatory of Japan and University of Tokyo
in collaboration with South African Astronomical Observatory under
a financial support of Grant-in-Aid for Scientific Research
on Priority Area (A) No. 10147207 and 10147214 of the Ministry of Education,
Culture, Sports, Science and Technology of Japan.
This work was supported by Grant-in-Aid for Scientific Research
(No.~15071204, 15340061, 19204018, 21540240 and 07J05097).
In addition,
NM acknowledges the support by Grant-in-Aid for Research Activity Start-up
(No.~22840008) and Grant-in-Aid for Young Scientists (No.~80580208)
from the Japan Society for the Promotion of Science (JSPS).
MWF gratefully acknowledges the receipt of a research grant from
the national Research Council of South Africa (NRF).
This publication makes use of data products from the Two Micron All Sky Survey,
which is a joint project of the University of Massachusetts and
the Infrared Processing and Analysis Center/California Institute of Technology,
funded by the National Aeronautics and Space Administration and
the National Science Foundation.

\bibliographystyle{mn2e}

\appendix

\section*{Supporting information}

Additional Supporting Information may be found
in the online version of this article. \\
{\bf Table \ref{tab:LCdata}.}~The released table of light variation for the catalogued variables. \\
{\bf Table \ref{tab:OGLE2MASS}.}~2MASS counterparts for the OGLE-III type II Cepheids in the bulge.

\label{lastpage}

\end{document}